\documentclass[%
 aip,
 jmp,%
 amsmath,amssymb,
]{revtex4-1}

\usepackage{graphicx}
\usepackage{dcolumn}
\usepackage{bm}
\usepackage{subfigure} 
\usepackage[dvipdfm, usenames]{color}

\begin{document}


\title[Line tension in cavity]{Line tension and morphology of a droplet and a bubble attached to the inner wall of a spherical cavity
}

\author{Masao Iwamatsu}
\email{iwamatsu@ph.ns.tcu.ac.jp}
\affiliation{Department of Physics, Faculty of Liberal Arts and Sciences, Tokyo City University, Setagaya-ku, Tokyo 158-8557, Japan.}
\date{\today}

\begin{abstract}
The effects of line tension on the morphology of a lens-shaped droplet and bubble placed on the inner wall of a spherical cavity are studied.  The contact angle between the lens-shaped droplet and the concave spherical substrate is expressed by the generalized Young's formula.  The equator of the spherical substrate is found to play a crucial role.  Neither a droplet with its contact line on the upper hemisphere of the substrate nor one with its contact line on the lower hemisphere can transform into each other continuously.  On a hydrophobic substrate, the contact angle jumps discontinuously to $180^{\circ}$, and the droplet is detached from the substrate to form a spherical droplet when the line tension is positive and large. This is similar to the drying transition on a flat substrate.  On the other hand,  on a hydrophilic substrate, the contact angle jumps discontinuously to $0^{\circ}$ when the line tension is positive and large.  Then, the droplet spread over the whole inner wall to leave a spherical bubble.  Therefore, not only the drying transition but also the wetting transition is induced by positive line tension on a concave spherical substrate.  There also exist stable as well as metastable droplets, whose phase diagrams can be complex.  When the line tension is negative and its magnitude increases, the contact line approaches the equator infinitesimally from either above or below.  However, it cannot cross the equator of a spherical cavity continuously.  The droplet with a contact line that coincides with the equator is a singular droplet. The contact line is pinned and cannot move, irrespective of the magnitude of the line tension.
\end{abstract}

\maketitle

\section{\label{sec:sce1}Introduction}
An understanding of the contact angle of micro- and nano-droplets and bubbles is an urgent necessity, as it is related to the analysis and development of various micro- and nano-devices based on droplets and bubbles~\cite{Seemann2012,Lohse2015}.  In particular, the wetting and drying strategies borrowed from biological structures have potential for the development and design of new materials following the design principle known as biomimetics~\cite{Nosonovsky2007,Song2014}.  When a droplet wets a substrate, the line tension~\cite{Gibbs1906,deGennes1985,Bonn2009,Weijs2011,Bormashenko2013} at the three-phase contact line should play some role in determining the morphology of the droplet.  The line tension is particularly important for nanoscale droplets, as the magnitude of the line tension is quite small~\cite{Pompe2000,Wang2001,Checco2003,Schimmele2007,Bonn2009}.  It has already been pointed out that the line tension plays a fundamental role in the stability of a non volatile droplet~\cite{Widom1995} and in the heterogeneous nucleation of a volatile droplet~\cite{Navascues1981,Singha2015} on a flat substrate.

However, the line-tension effect has been primarily considered on a flat substrate~\cite{Navascues1981,Widom1995,Blecua2006,Singha2015}.  There have been almost no theoretical or experimental attempts to clarify the line-tension effects on various substrates with complex geometries except for a very small number of studies done on the line-tension effect on a {\it convex} spherical substrate~\cite{Guzzardi2007,Hienola2007,Cooper2007,Dutka2010,Iwamatsu2015a,Iwamatsu2015b,Qiu2015}. The number of studies done on {\it concave} substrates is even smaller~\cite{Kubalsky2000,Dubrovskii2009,Extrand2012,Maksimov2013,Maheshwari2016}, though they play important role in various phenomena such as heterogeneous nucleation~\cite{Ruckenstein2010,Qian2012}, wetting of a structured surface~\cite{Whyman2011} and so on.

In this paper, we will consider the relationship between the line tension and the morphology of a droplet and a bubble placed on an inner wall of a spherical cavity.  Since the volume of the cavity is fixed, the droplet and bubble can be studied on the same footing.  We will mainly consider the droplet of a nonvolatile liquid.  We extend our previous work~\cite{Iwamatsu2015a,Iwamatsu2015b} on  a volatile liquid and consider the line-tension effects on the morphology of a lens-shaped droplet of a non-volatile liquid placed on the bottom of a spherical cavity.  This problem is equivalent to that of a lens-shaped bubble attached to the top of the inner wall of a spherical cavity.  

We find again a special role played by the equator of the concave spherical cavity, which was found previously for a droplet placed on a convex spherical substrate~\cite{Iwamatsu2015a,Iwamatsu2015b}.  When the contact line is on the upper hemisphere, it cannot cross the equator continuously to move into the lower hemisphere by increasing the magnitude of the negative line tension, and vice versa. A droplet whose contact line coincides with the equator of the cavity is a special case.  The contact line is fixed at the equator and cannot move even if the magnitude of the positive and negative line tension is altered.  In contrast to a flat substrate where only the drying transition is expected for the positive line tension, we observe the wetting transition in addition to the drying transition~\cite{Widom1995} by increasing the positive line tension when the contact line does not coincide with the equator.  More specifically, the droplet will form a spherical droplet in the drying transition and it will spread over the whole wall of the inner substrate of the cavity to leave a spherical bubble in the wetting transition. In this paper, we will use the terminologies "wetting/drying transitions" although they are really the finite-size wetting/drying transitions because the size of liquid droplet is finite. In section II, we will formulate the line tension effect on the Helmholtz free energy and the contact angle derived from the minimization of the free energy.  In section III, we will discuss the scenario of the morphological transition outlined above using the mathematically rigorous formulation of section II.  We will conclude in section IV.

\section{\label{sec:sec2}Line-tension effects on the Helmholtz free energy in a cavity}

\begin{figure}[htbp]
\begin{center}
\includegraphics[width=0.90\linewidth]{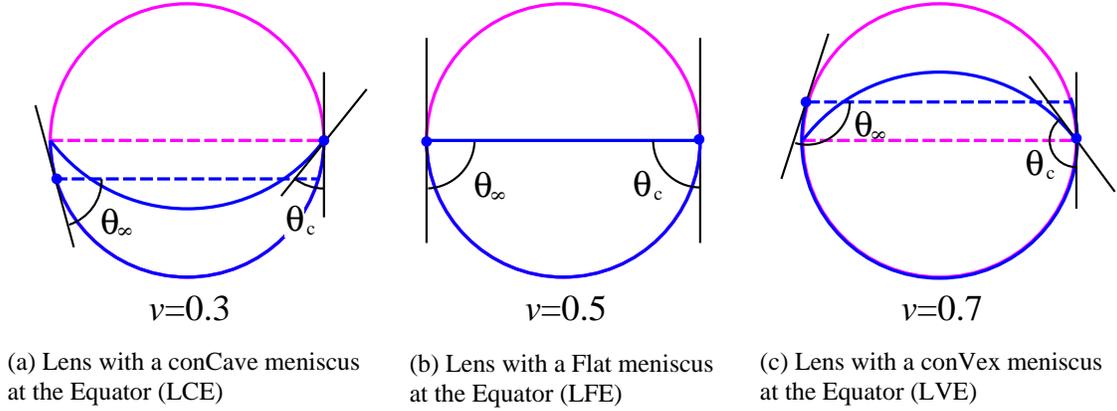}
\caption{
(a) A Lens-shaped droplet of contact angle $\theta_{\rm c}$ with concave meniscus ($\theta_{\rm c}<\theta_{\infty}$) when the contact line coincides with the equator (LCE).  The contact angle $\theta_{\infty}$ represents the contact angle for a flat meniscus.  Note that this droplet with concave meniscus is equivalent to a lens-shaped bubble with a convex meniscus. (b) A droplet with a flat meniscus  ($\theta_{\rm c}=\theta_{\infty}$) when the contact line coincides with the equator (LFE).  (c) A droplet with a convex meniscus ($\theta_{\rm c}>\theta_{\infty}$) when the contact line coincides with the equator (LVE).  The contact angle $\theta_{\rm c}$ is the characteristic contact angle defined by Eq.~(\ref{eq:C18}).  The contact angle $\theta_{\infty}$ for a flat meniscus is determined from Eq.~(\ref{eq:C1}). When the volume of the droplet is just half of the volume of cavity ($v=0.5$), $\theta_{\rm c}=\theta_{\infty}=90^{\circ}$ and the contact line coincides with the equator as shown in (b).
   }
\label{fig:CC1}
\end{center}
\end{figure}

In our previous work~\cite{Iwamatsu2015a,Iwamatsu2015b}, we considered the line-tension effects on a critical droplet of a volatile liquid heterogeneously nucleated on convex and concave spherical substrates and calculated the Gibbs free energy, which is appropriate for studying the activation energy of nucleation.  In this study, we focus on the physics of the line tension on a lens-shaped droplet of a non-volatile liquid placed on a spherical concave substrate of a spherical cavity, as shown in Fig.~\ref{fig:CC1}.  We consider a droplet of a non-volatile liquid with radius $r$ and contact angle $\theta$, placed on the bottom of the inner wall of a spherical cavity of radius $R$.  The droplet volume  $V$ is held constant and, therefore, the radius $r$ depends on the contact angle $\theta$.   We use the so-called {\it capillary model}, where the structure and width of the interfaces are neglected and the liquid-vapor, liquid-sold, and solid-vapor interactions are accounted for by the curvature-independent surface tensions. 

Although any axial-symmetric droplet with non-spherical surface will be possible~\cite{Dutka2010}, in particular, when the liquid-substrate molecular interaction or the disjoining pressure is important~\cite{Bormashenko2013,Pompe2000,Wang2001,Checco2003}, we will concentrate on the spherical lens-shaped droplet in this work for the first step towards the understanding of the line-tension effect on the droplet in cavity.  Once the molecular interaction becomes important and cannot be neglected, the line tension is not a constant but depends on the disjoining pressure~\cite{Bormashenko2013,Pompe2000,Wang2001,Checco2003} and the geometries of droplet and substrate.   It is also well known that the surface tension depends on the curvature of the surface~\cite{Tolman1949}.  For example, the liquid-vapor surface tension $\sigma_{\rm lv}$ depend on the radius $r$ of the surface as 
$\sigma_{\rm lv}\left(r\right)=\sigma_{\rm lv}\left(r\rightarrow\infty\right)/\left(1+2\delta/r\right)$, where $\delta$ is the first order correction term known as the "Tolman length".  Although Tolman~\cite{Tolman1949} himself suggested that $\delta$ is of the order of a molecular diameter and positive for droplet, the subsequent works~\cite{Block2014} revealed that $\delta$ is much smaller than the molecular diameter.  Therefore, the curvature dependence of the surface tension can be neglected as far as we consider a droplet of nano- and micro-scale size. As has been noted in the previous publication~\cite{Iwamatsu2015b}, the line tension must also be curvature dependent; otherwise, the force balance condition will be unphysical for a small droplet with $r\rightarrow 0$.  Again, we will neglect the curvature-dependence of line tension since we consider a droplet with a finite size.

The meniscus of a droplet can be convex, concave or flat depending on the magnitude of the contact angle $\theta$.  The contact angle $\theta_{\infty}$ for a flat substrate is determined from the implicit equation
\begin{equation}
V=\frac{\pi}{3}\left(2-3\cos\theta_{\infty}+\cos^{3}\theta_{\infty} \right)r^{3}
\label{eq:C1}
\end{equation}
for the droplet volume $V$.  In Fig.~\ref{fig:CC1}, we show the three types of meniscus when the contact line coincides with the equator of the cavity.  In this case, the contact angle $\theta=\theta_{\rm c}$ is fixed and not affected by the presence of line tension, which will be discussed later.  If $\theta_{\rm c}<\theta_{\infty}$, the meniscus becomes concave (Fig.~\ref{fig:CC1}(a), $v=0.3$).  However, if $\theta_{\rm c}>\theta_{\infty}$, the meniscus becomes convex (Fig.~\ref{fig:CC1}(c), $v=0.7$).  If $\theta_{\rm c}=\theta_{\infty}$ the meniscus is flat  (Fig.~\ref{fig:CC1}(b), $v=0.5$).

\begin{figure}[htbp]
\begin{center}
\includegraphics[width=0.70\linewidth]{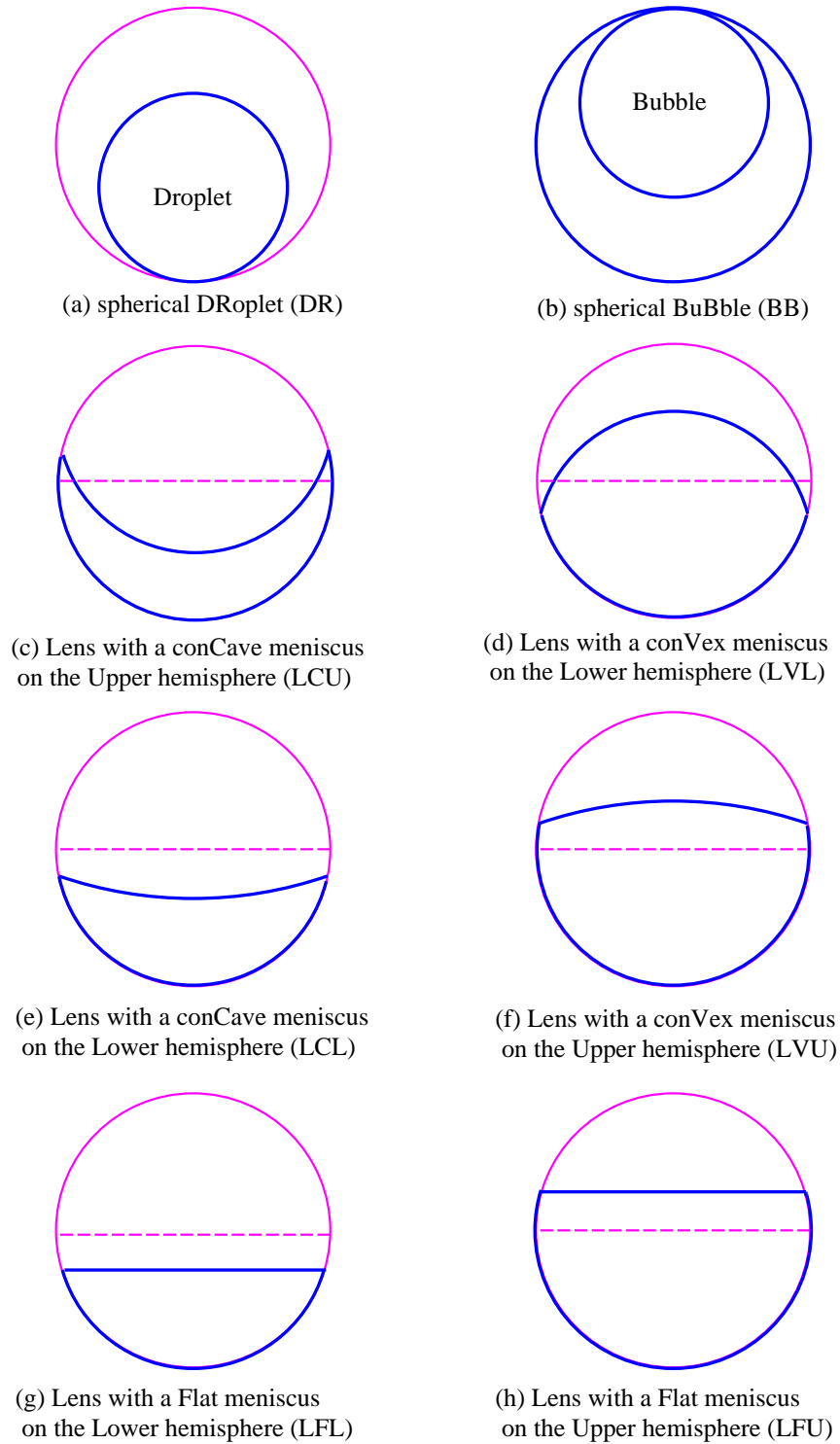}
\caption{
A catalog of droplet morphologies.  The droplet volume of the left column is $v=0.3$ and that of the right column is $v=0.7$.  The droplet attached to the bottom of the cavity on the left and the bubble attached to the top of the cavity on the right have the same shape.
 }
\label{fig:CC2}
\end{center}
\end{figure}

In addition to these three types of droplet, whose contact lines coincide with the equator as shown in Fig.~\ref{fig:CC1}, we expect eight different types of droplet morphology, as shown in Fig.~\ref{fig:CC2}:
\renewcommand{\labelenumi}{(\alph{enumi})}
\begin{enumerate}
\item A spherical droplet with $\theta=180^{\circ}$ (DR).  As a result, complete drying of the concave substrate will occur.
\item A spherical bubble with $\theta=0^{\circ}$ (BB).   As a result, complete wetting of the concave substrate will occur.
\item A lens-shaped droplet with a concave meniscus ($\theta<\theta_{\infty}$) whose contact line is on the upper hemisphere (LCU).
\item A lens-shaped droplet with a convex meniscus ($\theta>\theta_{\infty}$) whose contact line is on the lower hemisphere (LVL).
\item A lens-shaped droplet with a concave meniscus ($\theta<\theta_{\infty}$) whose contact line is on the lower hemisphere (LCL).
\item A lens-shaped droplet with a convex ($\theta>\theta_{\infty}$) meniscus whose contact line is on the upper hemisphere (LVU).
\item A lens-shaped droplet with a flat meniscus whose contact line is on the lower hemisphere (LFL).
\item A lens-shaped droplet with a flat meniscus whose contact line is on the upper hemisphere (LFU).
\end{enumerate}
Apparently, the droplet with a flat meniscus is the boundary case between a concave and a convex meniscus.  Therefore, LFL is the boundary of LCL and LVL, and LFU is the boundary of LCU and LVU (Fig.~\ref{fig:CC2}).  Note that the morphology of a droplet (bubble) on the left column of Fig.~\ref{fig:CC2} is the same as that of a bubble (droplet) on the right column.  The purpose of this paper is to study the transformation of the droplet (bubble) morphologies, listed in Fig.~\ref{fig:CC2} induced by the action of line tension.  We should note in passing that this droplet-bubble symmetry is valid only when the long-range molecular force known as the disjoining pressure~\cite{Bormashenko2013,Pompe2000,Wang2001,Checco2003}, which must be different for droplets and bubbles, can be neglected (capillary model).  We have to study the droplet and the bubble separately when the molecular forces cannot be represented by simple surface tensions.

In order to determine the most stable droplet shape, we have to identify the morphology which minimizes the Helmholtz free energy of a droplet in the capillary model given by
\begin{equation}
F=\sigma_{\rm lv}A_{\rm lv}+\Delta\sigma A_{\rm sl}+\tau L,
\label{eq:C2}
\end{equation}
and
\begin{equation}
\Delta\sigma = \sigma_{\rm sl}-\sigma_{\rm sv}=-\sigma_{\rm lv}\cos\theta_{\rm Y},
\label{eq:C3}
\end{equation}
where $A_{\rm lv}$ and $A_{\rm sl}$ are the surface areas of the liquid-vapor and liquid-solid (substrate) interfaces, respectively, and $\sigma_{\rm lv}$ and $\sigma_{\rm sl}$ are their respective surface tensions.  Moreover, $\Delta \sigma$ is the free energy gained when the solid-vapor interface with surface tension $\sigma_{\rm sv}$ is replaced by the solid-liquid interface with surface tension $\sigma_{\rm sl}$.   This free energy gain $\Delta\sigma$ is characterized by the wettability (hydrophilicity and hydrophobicity) of the substrate represented by Young's contact angle $\theta_{\rm Y}$.  Eq.~(\ref{eq:C3}) is known as the classical Young's equation~\cite{Young1805}.  The effect of the line tension $\tau$ is given by the last term in Eq.~(\ref{eq:C2}), where $L$ denotes the length of the three-phase contact line.  When the line tension is positive ($\tau>0$), the droplet tends to minimize or even vanish the line length  $L$ to lower the free energy $F$.  When the line tension is negative ($\tau<0$), the droplet tends to maximize the line length  $L$.

The contact angle $\theta$ is determined by minimizing the Helmholtz free energy Eq.~(\ref{eq:C2}) with respect to the radius $r$ of the droplet under the condition of a constant volume given by
\begin{equation}
V=\frac{4\pi}{3}R^{3}\omega\left(\rho,\theta\right),
\label{eq:C4}
\end{equation}
with
\begin{eqnarray}
\omega\left(\rho,\theta\right)&=&\frac{1}{16\xi}\left(\xi-1-\rho\right)^{2} \nonumber \\
&&\times\left[3\left(1-\rho\right)^{2}-2\xi\left(1+\rho\right)-\xi^{2}\right],\;\;\;  \nonumber \\
&&(\theta>\theta_{\infty}, \mbox{Convex}),
\label{eq:C5}
\end{eqnarray}
for the convex meniscus where
\begin{equation}
\xi=\sqrt{1+\rho^{2}+2\rho\cos\theta},
\label{eq:C6}
\end{equation}
and
\begin{equation}
\rho=\frac{r}{R}
\label{eq:C7}
\end{equation}
is the size parameter of the droplet.  Note that the radius $r$ and the size parameter $\rho=\rho\left(\theta\right)$ are functions of the contact angle $\theta$ as the volume $V$ of the droplet is fixed.  Equation (\ref{eq:C5}) was derived using the integration scheme originally developed by Hamaker~\cite{Hamaker1937}.  The detailed derivation of the volume as well as the Helmholtz energy was detailed in our previous paper~\cite{Iwamatsu2015a} and given in the Appendix.

Similarly, the droplet volume for the concave meniscus is given by
\begin{eqnarray}
\omega\left(\rho,\theta\right)&=&\frac{1}{16\zeta}\left(\zeta+1-\rho\right)^{2} \nonumber \\
&&\times\left[3\left(1+\rho\right)^{2}+2\zeta\left(1-\rho\right)-\zeta^{2}\right], \nonumber \\
&&(\theta<\theta_{\infty}, \mbox{Concave}),
\label{eq:C8}
\end{eqnarray}
where
\begin{equation}
\zeta=\sqrt{1+\rho^{2}-2\rho\cos\theta}.
\label{eq:C9}
\end{equation}
The result (Eq.~(\ref{eq:C8})) for the concave meniscus can be derived simply by changing the sign of $\rho$ and $\xi$ in Eq.~(\ref{eq:C5}) as follows
\begin{equation}
\rho \rightarrow -\rho,\;\;\;\; \xi \rightarrow -\zeta.
\label{eq:C10}
\end{equation}
Therefore, we will only present the formulae for the convex meniscus from now on for brevity.  The formulae for the concave meniscus can be easily derived using the transformation in Eq.~(\ref{eq:C10}).

Within the capillary approximation, the Helmholtz free energy Eq.~(\ref{eq:C2}) is given by
\begin{equation}
F=4\pi R^{2}\sigma_{\rm lv}f\left(\rho,\theta\right),
\label{eq:C11}
\end{equation}
with
\begin{eqnarray}
f\left(\rho,\theta\right)
&=& \rho\frac{1-\left(\xi-\rho\right)^{2}}{4\xi}-\cos\theta_{\rm Y}\frac{\rho^{2}-\left(\xi-1\right)^{2}}{4\xi}+\frac{\tilde{\tau}\rho}{2\xi}\sin\theta, \nonumber \\
&&\;\;\;\;\;\;\;\;(\theta>\theta_{\infty}, \mbox{Convex}),
\label{eq:C12}
\end{eqnarray}
where
\begin{equation}
\tilde{\tau}=\frac{\tau}{\sigma_{\rm lv}R}
\label{eq:C13}
\end{equation}
is the scaled line tension. The free energy for a concave meniscus will be obtained using the transformation in Eq.~(\ref{eq:C10}).  Since the Helmholtz free energy in Eq.~(\ref{eq:C2}) is proportional to the surface area of a droplet, the free energy will increase if a single droplet breaks up into a multiple droplets.  Therefore, we will not consider the situation of a multiple cap-shaped droplet covering a single spherical substrate.

Minimization of the Helmholtz free energy under the constant volume constraint leads to an equation that determines the equilibrium contact angle $\theta_{\rm e}$ and the radius $\rho_{\rm e}=\rho\left(\theta_{\rm e}\right)$ written as
\begin{eqnarray}
&&\left(\cos\theta_{\rm Y}-\cos\theta_{\rm e}\right)-\tilde{\tau}\frac{1+\rho_{\rm e}\cos\theta_{\rm e}}{\rho_{\rm e}\sin\theta_{\rm e}}=0, \nonumber \\
&&\;\;\;\;\;\;\;\;\;(\theta_{\rm e}>\theta_{\infty}, \mbox{Convex}),
\label{eq:C14}
\end{eqnarray}
for a lens-shaped droplet with a convex meniscus~\cite{Iwamatsu2015a,Iwamatsu2015b}.  The corresponding minimized (extremized) free energy of a lens-shaped droplet is given by
\begin{equation}
F_{\rm lens}=4\pi R^{2}\sigma_{\rm lv}f_{\rm lens},
\label{eq:C15}
\end{equation}
with
\begin{eqnarray}
f_{\rm lens}&=&\frac{\left(1+\rho_{\rm e}-\xi_{\rm e}\right)^{2}\left(\cos\theta_{\rm e}+1+\xi_{\rm e}\right)}{4\xi_{\rm e}}
\nonumber \\
&&-\tilde{\tau}\frac{\left(1+\rho_{\rm e}\cos\theta_{\rm e}-\xi_{\rm e}\right)}{2\rho_{\rm e}\sin\theta_{\rm e}},\nonumber \\
&&(\theta_{\rm e}>\theta_{\infty}, \mbox{Convex}),
\label{eq:C16}
\end{eqnarray}
for the convex meniscus, where $\xi_{\rm e}$ is given by Eq.~(\ref{eq:C6}) with $\rho$ and $\theta$ replaced by $\rho_{\rm e}=\rho\left(\theta_{\rm e}\right)$ and $\theta_{\rm e}$ determined from Eq.~(\ref{eq:C14}).  Eq.~(\ref{eq:C16}) reduces to the well-known formula~\cite{Navascues1981} for the Helmholtz free energy of a droplet on a flat substrate when $\rho\rightarrow 0$ or $R\rightarrow\infty$.  

The results which correspond to Eqs.~(\ref{eq:C14}) and (\ref{eq:C16}) for a concave meniscus are obtained by using the transformation  in Eq.~(\ref{eq:C10}).   Eq.~(\ref{eq:C14}) is similar to the classical Young's equation~\cite{Young1805} on a flat substrate given by Eq.~(\ref{eq:C3}).  In fact, even on a concave spherical substrate, the contact angle is determined from the classical Young's equation (\ref{eq:C3}) for flat surfaces~\cite{Qian2012}, and $\theta_{\rm e}=\theta_{\rm Y}$ from Eq.~(\ref{eq:C14}) when the line tension can be neglected ($\tau=0$).

All the solutions of Eq.~(\ref{eq:C14}) do not necessarily correspond to those for the stable lens-shaped droplet.  In fact, some of them may correspond to the maximum rather than the minimum of the Helmholtz free energy.  It is possible to determine the stability limit of a droplet by calculating the second derivative of the Helmholtz free energy in Eq.~(\ref{eq:C2}).  The detailed derivation is rather lengthy and the result are given in the Appendix.  This stability limit is similar to the spinodal of a first-order phase transition.  The stability condition of the lens-shaped droplet is given by
\begin{eqnarray}
\tilde{\tau} &\le& \frac{\rho_{\rm e}\left(\rho_{\rm e}+\cos\theta_{\rm e}+2\sqrt{1+\rho_{\rm e}^{2}+2\rho_{\rm e}\cos\theta_{\rm e}}\right)\sin^{3}\theta_{\rm e}}{1+\rho_{\rm e}^{2}+2\rho_{\rm e}\cos\theta_{\rm e}} \nonumber \\
&\equiv& \tilde{\tau}_{\rm st}\left(\theta_{\rm e}\right), \nonumber \\ 
&&\;\;\;\;\;\;\;\;\;\;\;\;\;\;\;\;(\theta_{\rm e}>\theta_{\infty}, \mbox{Convex}).
\label{eq:C17}
\end{eqnarray}
When the line tension is larger than the stability limit $\tilde{\tau}_{\rm st}$ ($\tilde{\tau}>\tilde{\tau}_{\rm st}$), the lens-shaped droplet is unstable and cannot form on a concave spherical substrate, as the Helmholtz free energy given by Eq.~(\ref{eq:C16}) is maximum rather than minimum.  Then the droplet may form a spherical droplet, as shown in Fig.~\ref{fig:CC2}(a), or it may spread over the spherical substrate to form a spherical bubble as shown in Figs.~\ref{fig:CC2}(b). The morphology of the droplet will be determined from the free energy in Eq.~(\ref{eq:C16}) and the stability condition in Eq.~(\ref{eq:C17}), as will be discussed in section III.

As has been pointed out by Hienola {\it et al.}~\cite{Hienola2007},  the generalized Young's equation in Eq.~(\ref{eq:C14}) can also be derived from the mechanical-force balance of the surface tensions $\sigma_{\rm lv}$, $\sigma_{\rm sv}$, $\sigma_{\rm sl}$, and the line tension~\cite{Iwamatsu2015b}.   The line tension $\tau$ does not contribute to the force balance or, therefore, to the determination of contact angle when the contact line coincides with the equator of the spherical cavity, though it does contribute to the free energy~\cite{Iwamatsu2015a,Iwamatsu2015b}.  Then, the contact angle is given by the characteristic contact angle $\theta_{\rm c}$ determined from
\begin{equation}
1+\rho\cos\theta_{\rm c}=0,\;\;\;(\theta_{\rm c}>\theta_{\infty}, \mbox{Convex}),
\label{eq:C18}
\end{equation}
for the convex meniscus from Eq.~(\ref{eq:C14}).  Fig. \ref{fig:CC1} showed the three types of droplet whose contact line coincides with the equator.  In this case, the contact line is pinned at the equator even if the magnitude of the line tension is altered because the line tension cannot affect the force balance~\cite{Iwamatsu2015b}.  Physically, the contact line cannot move as the movement of the circular contact line either toward the upper hemisphere or toward the lower hemisphere leads to a {\it decrease} of the contact line length.  The contact line length is maximum at the equator if it has a circular shape.  The equator plays the role of infinity of the infinite flat substrate~\cite{Iwamatsu2015b}.

Because the size parameter $\rho=\rho\left(\theta\right)$ is a function of $\theta$ from Eq.~(\ref{eq:C5}) for a fixed volume $V$ or a fixed scaled volume $v=V/(4\pi R^3/3)$, the size parameter $\rho_{\rm e}=\rho\left(\theta_{\rm e}\right)$ in the generalized Young's equation Eq.~(\ref{eq:C14}) is also a function of the equilibrium contact angle $\theta_{\rm e}$. In order to fix the droplet volume, it will be convenient to characterize the droplet volume when it is spherical with the contact angle $\theta=180^{\circ}$ and the radius $\rho_{\pi}=\rho\left(\theta=180^{\circ}\right)$. Then, the size parameter $\rho$ is a function of the contact angle $\theta$ through the implicit equation:
\begin{equation}
\omega\left(\rho,\theta\right)=\rho_{\pi}^{3}=v,\;\;\;\rightarrow\;\; \rho=\rho\left(\theta\right).
\label{eq:C19}
\end{equation}
Naturally, the cavity cannot accommodate a droplet whose volume is larger than the cavity volume.  Therefore, $0\leq v \leq 1$ and $0\leq \rho_{\pi} \leq 1$.

\begin{figure}[htbp]
\begin{center}
\includegraphics[width=0.70\linewidth]{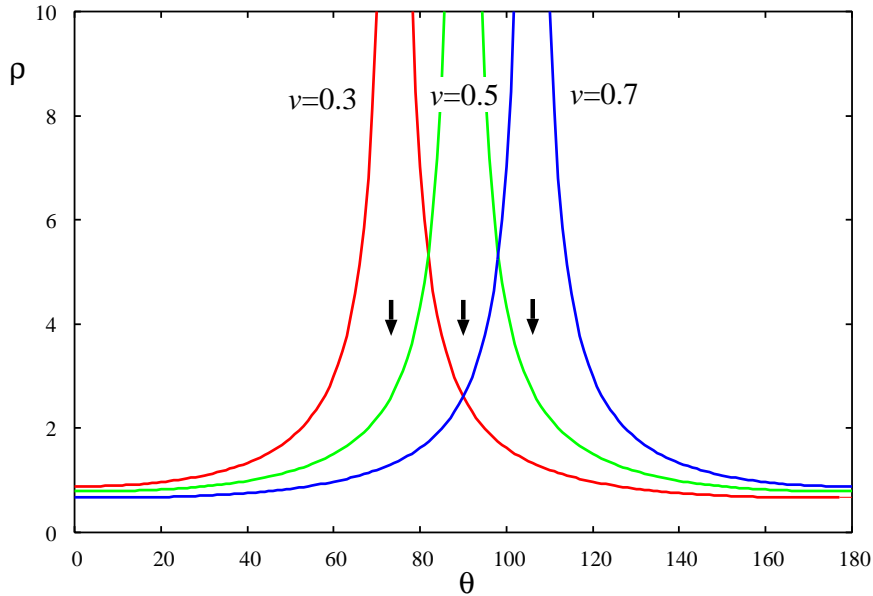}
\caption{
The size parameter $\rho\left(\theta\right)$ as a function of the contact angle $\theta$ for $v=0.3$, $v=0.5$ and $0.7$.  The radius $\rho$ diverges at $\theta_{\infty}$ indicated by an arrow determined from Eq.~(\ref{eq:C1}).
}
\label{fig:CC3}
\end{center}
\end{figure}

Figure \ref{fig:CC3} shows the size parameter $\rho\left(\theta\right)$ as a function of the contact angle $\theta$ for given volume $v$.  The size parameter $\rho$ and, therefore, the radius $r$ of the droplet increase from $\rho_{\pi}$ at $\theta=180^{\circ}$ as the contact angle is decreased.  The size parameter diverges ($\rho\rightarrow \infty$) at the contact angle $\theta_{\infty}$ determined from Eq.~(\ref{eq:C1}) where the meniscus becomes flat and it changes from concave to convex.  The three-phase contact line coincides with the equator when the contact angle becomes the characteristic contact angle ($\theta=\theta_{\rm c}$) determined from Eq.~(\ref{eq:C18}), where the size parameter remains finite and changes continuously.

The free energy $F_{\rm drop}$ of a spherical droplet DR (Fig.~\ref{fig:CC2}(a)) is given by the limit $\theta\rightarrow 180^{\circ}$ of Eq.~(\ref{eq:C12}), and is written as
\begin{equation}
F_{\rm drop}=4\pi R^{2}\sigma_{\rm lv}f_{\rm drop},
\label{eq:C20}
\end{equation}
where
\begin{equation}
f_{\rm drop}=\left(\rho_{\pi}\right)^{2}=\left(v\right)^{2/3}.
\label{eq:C21}
\end{equation}
If the Helmholtz free energy $f_{\rm lens}$ of the lens shaped droplet is higher than the free energy $f_{\rm drop}$ of the spherical droplet, the lens-shaped droplet will  transform into a spherical shape. Therefore, by comparing the free energy $f_{\rm lens}$ of the lens-shaped droplet with $f_{\rm drop}$ of a spherical droplet, we can study the morphological transition between a lens-shaped droplet and a spherical droplet, which is the {\it drying transition} predicted on a flat substrate~\cite{Widom1995}.

It is also possible to calculate the free energy $F_{\rm bubble}$ of a droplet that completely spread over the whole surface of a spherical cavity (Fig.~\ref{fig:CC2}(b)).  Then,  a spherical bubble (BB) attached to the top wall of the cavity will appear, which is realized when $\theta=0$.  The free energy is given by the $\theta\rightarrow 0^{\circ}$ limit of the free energy in Eq.~(\ref{eq:C12}) for the concave meniscus, which is given by
\begin{equation}
F_{\rm bubble}=4\pi R^{2}\sigma_{\rm lv}f_{\rm bubble},
\label{eq:C22}
\end{equation}
where
\begin{equation}
f_{\rm bubble}=\rho_{0}^{2}-\cos\theta_{\rm Y},
\label{eq:C23}
\end{equation}
and $\rho_{0}=\rho\left(\theta=0^{\circ}\right)$ is the size parameter when the contact angle is $\theta=0^{\circ}$.  By comparing the free energy $f_{\rm lens}$ of the lens-shaped droplet with $f_{\rm bubble}$ of the spherical bubble, we can study the morphological transition between a lens-shaped droplet and a spherical bubble, which might be termed the {\it wetting transition}.  

From the geometrical constraint that the volume of the droplet is fixed, we have
\begin{equation}
\left(\rho_{\pi}\right)^{3}=1-\left(\rho_{0}\right)^{3}.
\label{eq:C24}
\end{equation}
Then the relative stability of a spherical droplet and a spherical bubble is determined from $f_{\rm drop}=f_{\rm bubbe}$, which leads to the condition for the Young's contact angle
\begin{equation}
\theta_{\rm Y,w}=\cos^{-1}\left[\left(1-v\right)^{2/3}-\left(v\right)^{2/3}\right]
\label{eq:C25}
\end{equation}
where the wetting-drying transition occurs for a given droplet volume $v$.  A spherical droplet (Fig.~\ref{fig:CC2}(a)) is more stable than a spherical bubble (Fig.~\ref{fig:CC2}(b)) if $\theta_{\rm Y}>\theta_{\rm Y,w}$ and vice versa.  Young's contact angle $\theta_{\rm Y,w}$ plays the role of the boundary between hydrophobic and hydrophilic of the concave spherical substrate.  Importantly, the contact angle $\theta_{\rm Y,w}$ depends only on the droplet volume $v$ and can be controlled artificially.

\begin{figure}[htbp]
\begin{center}
\includegraphics[width=0.70\linewidth]{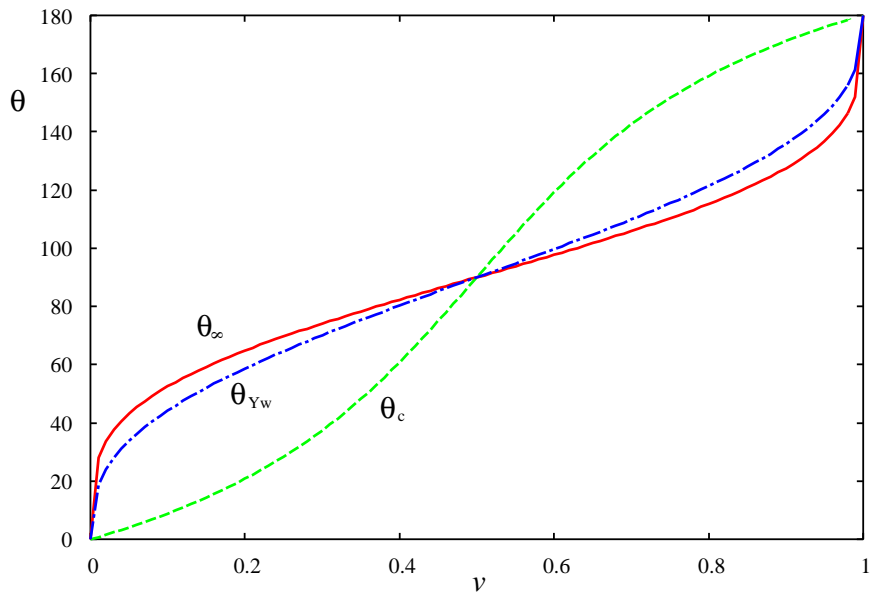}
\caption{
Young's contact angle $\theta_{\rm Y,w}$ given by Eq.~(\ref{eq:C25}) as a function of the droplet volume $v$, which separates a stable droplet and a stable bubble.  The contact angle $\theta_{\infty}$ for a flat meniscus calculated from Eq.~(\ref{eq:C1}) as well as the characteristic contact angle $\theta_{\rm c}$ from Eq.~(\ref{eq:C18}) for which the contact line coincides with the equator are also shown.   
  }
\label{fig:CC4}
\end{center}
\end{figure}

In Fig.~\ref{fig:CC4}, we show Young's contact angle $\theta_{\rm Y,w}$ of the neutral or hydrophilic-hydrophobic boundary determined from Eq.~(\ref{eq:C25}).  The contact angle $\theta_{\infty}$ for a flat meniscus determined from Eq.~(\ref{eq:C1}) and the characteristic contact angle $\theta_{\rm c}$ determined from Eq.~(\ref{eq:C18}) are also shown.  This hydrophilic-hydrophobic boundary is not necessarily $\theta_{\rm Y}=90^{\circ}$ but depends strongly on the droplet volume $v$.  When the droplet volume is a half of the cavity volume ($v=0.5$), all three contact angles coincide with $90^{\circ}$.  They all approach $0^{\circ}$ when $v\rightarrow 0$, and $180^{\circ}$ when $v\rightarrow 1$.

\section{\label{sec:sec3}  Morphological transition of a droplet on a concave spherical substrate}

\begin{figure}[htbp]
\begin{center}
\includegraphics[width=0.70\linewidth]{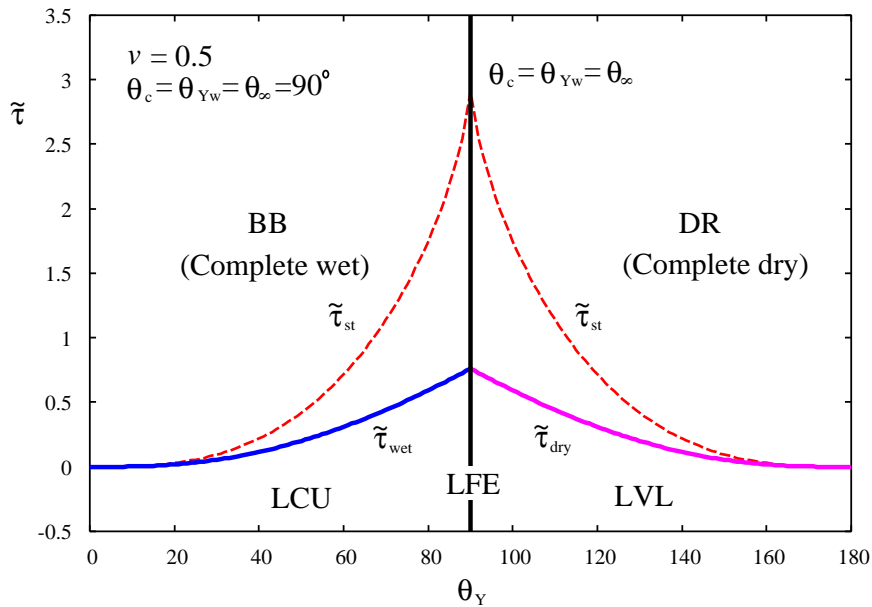}
\caption{
The morphological phase diagram of a droplet placed on the bottom of a spherical cavity when $v=0.5$ on the $\theta_{\rm Y}-\tilde{\tau}$ plane. In this case, the volume of the droplet is half of that of the cavity so that the phase diagram is symmetric between the bubble (BB) and the droplet (DR).
 }
\label{fig:CC5}
\end{center}
\end{figure}

By comparing the three free energies $f_{\rm lens}$, $f_{\rm drop}$ and $f_{\rm bubble}$, we can determine the most stable structure among the eleven morphologies in the catalog of Figs.~\ref{fig:CC1} and \ref{fig:CC2}.  Fig.~\ref{fig:CC5} shows the morphological phase diagram of a droplet placed on the bottom of the spherical cavity when $v=0.5$ on the $\theta_{\rm Y}-\tilde{\tau}$ plane.  In this case, $\theta_{\rm c}=\theta_{\rm Y,w}=\theta_{\infty}=90^{\circ}$, and the diagram is symmetrical about $\theta_{\rm Y}=90^{\circ}$.

The phase diagram in Fig.~\ref{fig:CC5} is divided into four regions by four boundaries indicated by two thick curves and one vertical tick line.  The dashed curves are the stability limit of the lens-shaped droplet.  Hence, only five morphologies, BB, DR, LCU, LFE and LVL (see Figs.~\ref{fig:CC1} and \ref{fig:CC2}) are stable.  The spherical bubble (BB) and the lens-shaped droplet with a concave meniscus whose contact line is on the upper hemisphere (LCU) can coexist along the curve indicated by $\tilde{\tau}_{\rm wet}$  (Fig.~\ref{fig:CC5}), which is determined from $f_{\rm bubble}=f_{\rm lens}$ together with Eqs.~(\ref{eq:C14}) and (\ref{eq:C19}) for the concave meniscus.  Similarly, the spherical droplet (DR) and the lens-shaped droplet with a convex meniscus whose contact line is on the {\color{Blue} lower} hemisphere (LVL) can coexist along the curve indicated by $\tilde{\tau}_{\rm dry}$ (Fig.~\ref{fig:CC5}), which is determined from $f_{\rm drop}=f_{\rm lens}$ together with Eqs.~(\ref{eq:C14}) and (\ref{eq:C19}) for a convex meniscus.  Finally, BB and DR, LCU and LVL are divided by the line $\theta_{\rm Y}=\theta_{\rm Y,w}=90^{\circ}$.  The boundary between LCU and LVL can be crossed continuously via the lens-shaped droplet LFE with a flat meniscus whose contact line coincides with the equator.

Above the two curves $\tilde{\tau}_{\rm wet}$ and $\tilde{\tau}_{\rm dry}$, which correspond to the wetting and drying transitions, the lens-shaped droplet (LCU and LVL) still can exist as a metastable droplet.  This metastable droplet becomes unstable above $\tilde{\tau}_{\rm st}$ (Fig.~\ref{fig:CC5}), which is determined from Eq.~(\ref{eq:C17}) together  with Eqs.~(\ref{eq:C14}) and (\ref{eq:C19}) for convex and concave menisci.  The stability limit $\tilde{\tau}_{\rm st}$ plays the role of the spinodal of the first-order phase transition.  

In order to observe this morphological transformation, the size of the scaled line tension is on the order of $\tilde{\tau}\simeq 0.1-1$.  Suppose the line tension is $\tau\sim 10^{-9}$ J/m~\cite{Pompe2000,Wang2001,Checco2003,Schimmele2007,Bonn2009}, the liquid-vapor surface tension is $\sigma_{\rm lv}\sim 70\times 10^{-3}$ J/m$^{2}$ (water), and the scaled line tension is $\tilde{\tau}\sim 1$, then the size of the cavity $R$ should be $R\sim 1\times 10^{-8}$ m=$10$ nm.  Therefore, submicron to nanometer size cavity will be necessary.  The magnitude of line tension which is necessary to observe this line-tension effect in a macro- and micro-scale cavity will be of the order of~\cite{Drelich1996}  $\tau\sim 10^{-5}-10^{-6}{\rm N}$.  However, an ultra law surface tension $\sigma_{\rm lv}\sim 10^{-7} {\rm N/m}$ and a line tension $\tau\sim 10^{-12}-10^{-13}{\rm N}$ have been predicted for a colloid-polymer mixture~\cite{Vandecan2008} recently.  Then, the size of the cavity can be  as large as $R \sim 10^{-5}-10^{-6}{\rm m}$, which could be easily observed by an optical microscope~\cite{Drelich1996}.

\begin{figure}[htbp]
\begin{center}
\subfigure[]
{
\includegraphics[width=0.45\linewidth]{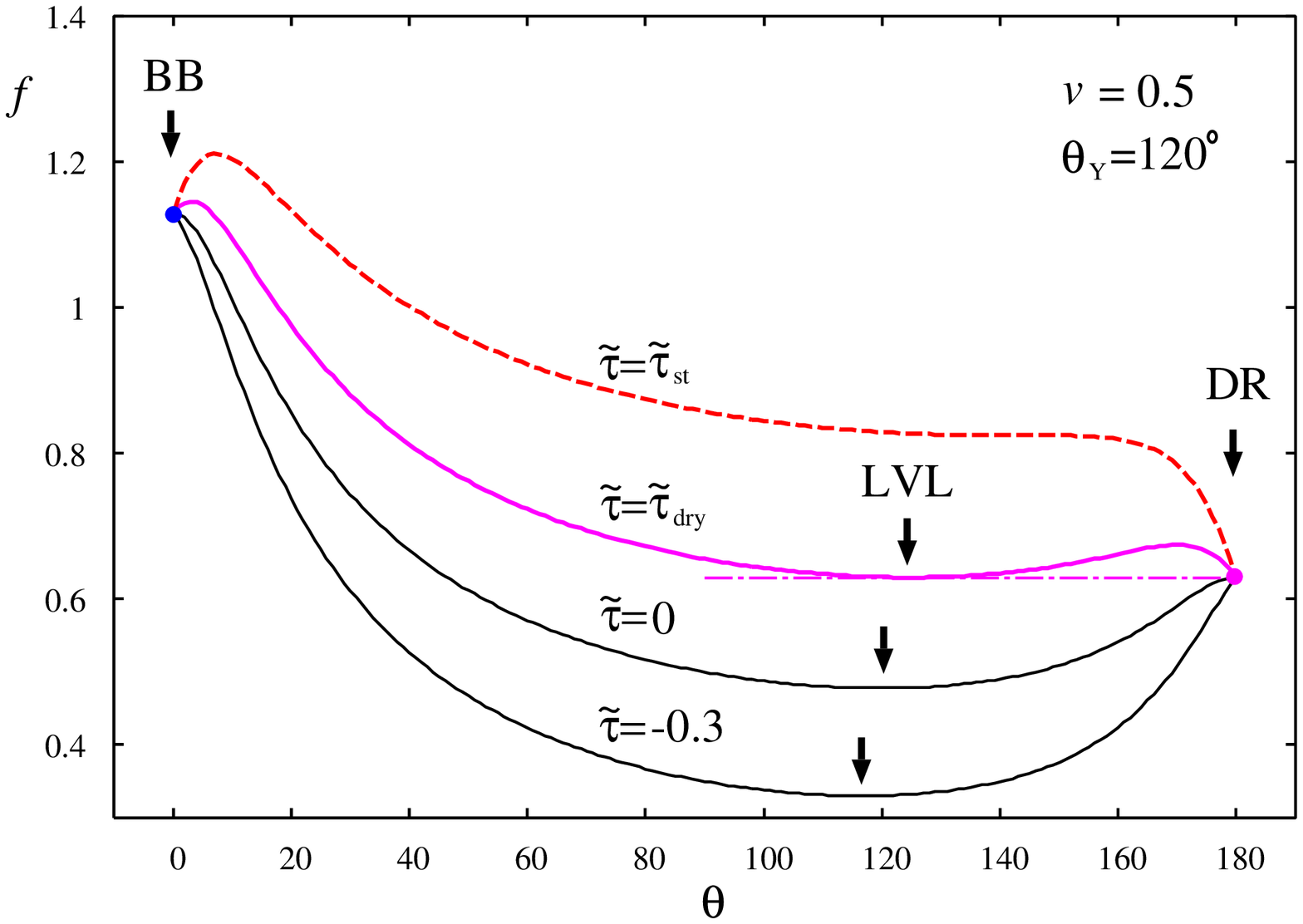}
\label{fig:6a}
}
\subfigure[]
{
\includegraphics[width=0.45\linewidth]{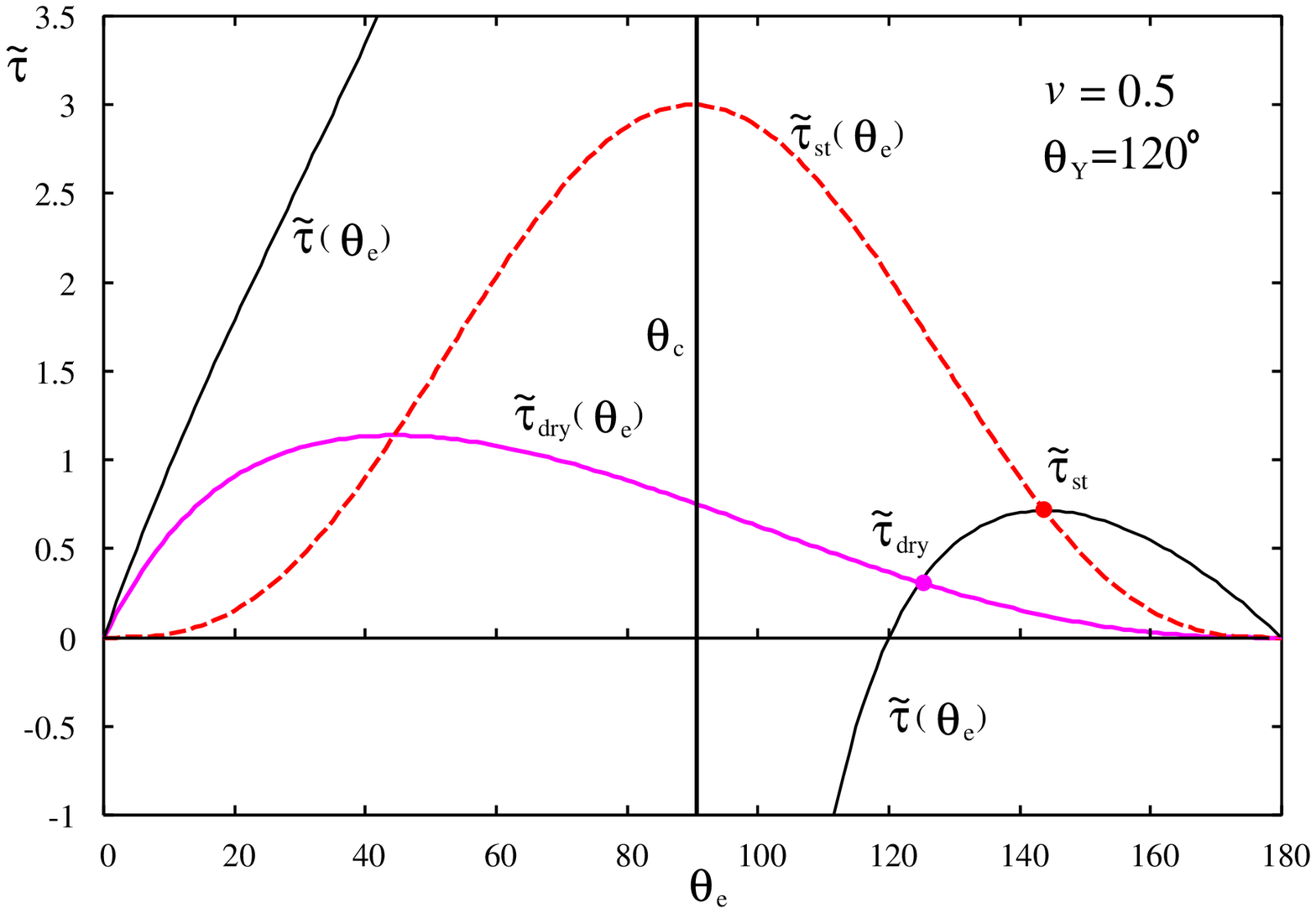}
\label{fig:6b}
}
\end{center}
\caption{
(a) The Helmholtz free energy landscape of Eq.~(\ref{eq:C12}) when $\theta_{\rm Y}=120^{\circ}$.  When $\tilde{\tau}=0$, the equilibrium contact angle $\theta_{\rm e}$ indicated by an arrow is given by the bare Young's contact angle $\theta=\theta_{\rm Y}=120^{\circ}$.  It increases as the line tension $\tilde{\tau}$ is increased.  When $\tilde{\tau}=\tilde{\tau}_{\rm dry}\simeq 0.310$, the first-order drying transition takes place: a lens-shaped droplet will transform into a spherical droplet.  When $\tilde{\tau}>\tilde{\tau}_{\rm st}\simeq 0.715$, a metastable lens-shaped droplet becomes unstable.  Then, only a stable droplet and a metastable bubble will exist. 
 (b) The determination of the equilibrium contact angle $\theta_{\rm e}$ as a function of the line tension $\tilde{\tau}$.  The transition points $\tilde{\tau}_{\rm dry}\simeq 0.310$ and $\tilde{\tau}_{\rm st}\simeq 0.715$ and their corresponding equilibrium contact angles are determined from the intersection of $\tilde{\tau}\left(\theta_{\rm e}\right)$, $\tilde{\tau}_{\rm dry}\left(\theta_{\rm e}\right)$ and $\tilde{\tau}_{\rm st}\left(\theta_{\rm e}\right)$.
 } 
\label{fig:CC6}
\end{figure}

Figure \ref{fig:CC6}(a) shows the Helmholtz free energy landscape for a hydrophobic wall with $\theta_{\rm Y}=120^{^\circ}$ calculated from Eqs~(\ref{eq:C12}) and (\ref{eq:C19}) as a function of the contact angle $\theta$ (see Fig.~\ref{fig:CC5}).   There are three free energy extrema, which correspond to a spherical bubble BB, a lens-shaped droplet LVL and a spherical droplet DR.  The free energy minimum for LVL is the lowest as far as $\tilde{\tau}<\tilde{\tau}_{\rm dry}\simeq 0.310$, where the two minima that correspond to LVL and DR have the same free energy so that they can coexist.  Then the drying transition of the cavity wall occurs.  When $\tilde{\tau}_{\rm st}>\tilde{\tau}>\tilde{\tau}_{\rm dry}$, the lens-shaped droplet LVL becomes metastable and the most stable morphology is a spherical droplet DR.  When $\tilde{\tau}>\tilde{\tau}_{\rm st}\simeq 0.715$, the lens shaped droplet LVL becomes unstable.  Then the most stable morphology is a spherical droplet DR, and a spherical bubble BB continues to be metastable.

From Eq.~(\ref{eq:C14}), we found
\begin{eqnarray}
\tilde{\tau}&=&\frac{\rho_{\rm e}\sin\theta_{\rm e}\left(\cos\theta_{\rm Y}-\cos\theta_{\rm e}\right)}{1+\rho_{\rm e}\cos\theta_{\rm e}}
\equiv \tilde{\tau}\left(\theta_{\rm e}\right), \nonumber \\
&&(\theta_{\rm e}>\theta_{\infty}, \mbox{Convex}).
\label{eq:C26}
\end{eqnarray}
The result for a concave meniscus will be obtained using the transformation in Eq.~(\ref{eq:C10}).  Fig.~\ref{fig:CC6}(b) shows $\tilde{\tau}\left(\theta_{\rm e}\right)$ when $\theta_{\rm Y}=120^{\circ}$ as a function of the equilibrium contact angle $\theta_{\rm e}$.   The curve consist of two curves which diverge at $\theta_{\rm e}=\theta_{\rm c}=90^{\circ}$ and a vertical line at $\theta_{\rm c}=90^{\circ}$.  The intersection of the horizontal line $\tilde{\tau}={\rm constant}$ and $\tilde{\tau}\left(\theta_{\rm e}\right)$ gives the equilibrium contact angle $\theta_{\rm e}$ for given $\tilde{\tau}$.  Apparently, $\theta_{\rm e}=\theta_{\rm Y}=120^{\circ}$ when $\tilde{\tau}\equiv 0$.  Also, $\theta_{\rm e}$ approaches $\theta_{\rm c}$ from above as the line tension $\tilde{\tau}$ becomes highly negative.  That is, the contact line approaches equator from below, and it remains on the lower hemisphere. 

In Fig.~\ref{fig:CC6}(b), we show $\tilde{\tau}_{\rm st}\left(\theta_{\rm e}\right)$ defined by Eq.~(\ref{eq:C17}).  Above this curve, a lens-shaped droplet is unstable.  The equilibrium contact angle $\theta_{\rm e}$ of a stable and a metastable droplet is larger than $\theta_{\rm c}$.  That is, neither a stable nor a metastable droplet whose contact line is on the upper hemisphere ($\theta_{\rm e}<\theta_{\rm c}$) can exist. The intersection of the curve $\tilde{\tau}_{\rm st}\left(\theta_{\rm e}\right)$ and $\tilde{\tau}\left(\theta_{\rm e}\right)$ gives the stability limit $\tilde{\tau}_{\rm st}\simeq 0.715$.  We also show the boundary of the drying transition $\tilde{\tau}_{\rm dry}\left(\theta_{\rm e}\right)$ from a lens-shaped droplet with convex meniscus LVL to the spherical droplet DR defined by
\begin{eqnarray}
\tilde{\tau}
&=&
\left(v^{2/3}-\frac{\left(1+\rho_{\rm e}-\xi_{\rm e}\right)^{2}\left(\cos\theta_{\rm e}+1+\xi_{\rm e}\right)}
{4\xi_{\rm e}}\right) \nonumber \\
&/&\left(\frac{\left(1+\rho_{\rm e}\cos\theta_{\rm e}-\xi_{\rm e}\right)}{2\rho_{\rm e}\sin\theta_{\rm e}}\right)\equiv \tilde{\tau}_{\rm dry}\left(\theta_{\rm e}\right), \nonumber \\
&&(\theta_{\rm e}>\theta_{\infty}, \mbox{Convex}),
\label{eq:C27}
\end{eqnarray}
which is derived from $f_{\rm lens}=f_{\rm drop}$ defined by Eqs.~(\ref{eq:C16}) and (\ref{eq:C21}).  Above $\tilde{\tau}_{\rm dry}\left(\theta_{\rm e}\right)$, a spherical droplet DR is most stable and a lens-shaped droplet LVL becomes metastable.  The intersection of the curves $\tilde{\tau}_{\rm dry}\left(\theta_{\rm e}\right)$ and $\tilde{\tau}\left(\theta_{\rm e}\right)$ gives the drying transition point $\tilde{\tau}_{\rm dry}\simeq 0.310$ for given $\theta_{\rm Y}=120^{\circ}$ in Fig.~\ref{fig:CC5}. 

The equilibrium contact angle $\theta_{\rm e}$, which is indicated by an arrow in Fig.~\ref{fig:CC6}(a), is determined from $\tilde{\tau}=\tilde{\tau}\left(\theta_{\rm e}\right)$ for given $\tilde{\tau}$.  As we increase the magnitude of the positive line tension $\tilde{\tau}$, the equilibrium contact angel $\theta_{\rm e}$ increases from $\theta_{\rm e}=\theta_{\rm Y}=120^{\circ}$ for $\tilde{\tau}=0$ along the line $\tilde{\tau}\left(\theta_{\rm e}\right)$ in Fig.~\ref{fig:CC6}(b). Apparently, the contact angle $\theta_{\rm e}$ does not increase continuously to $\theta_{\rm e}=180^{\circ}$ but it stops increasing at the stability limit $\tilde{\tau}_{\rm st}\simeq 0.715$. 

The effect of negative line tension on the droplet on a spherical substrate is different from that on a flat substrate.  As we increase the magnitude of the negative line tension ($\tilde{\tau}<0$), the contact angle of LVL infinitesimally approaches the characteristic contact angle from above ($\theta\rightarrow\theta_{\rm c}^{+}$).  In other words, the three-phase contact line indefinitely approaches the equator of the substrate from below.  The contact line remains in the lower hemisphere and cannot cross the equator.  This finding can be easily understood, because the contact-line length is maximized at the equator.  In order to maximize the negative gain of the line-tension contribution of the Helmholtz free energy, the contact line approaches the equator but never cross it.  Therefore, the contact line of the droplet always remains on the lower hemisphere irrespective of the magnitude of the line tension.

However, this result for the negative line tension is not conclusive as the undulation of the contact line necessarily increases the contact-line length and decreases the free energy further.  Then, a circular contact line might be unstable.  
Since we concentrate on the thermodynamics of the droplet, the stability of a lens-shaped droplet against the fluctuation which does not preserve the circular shape will not be considered.  It is well known that the capillary model of a lens-shaped droplet employed in this work has short-wavelength instability~\cite{Dobbs1999,Brinkmann2005,Guzzardi2006} on a flat substrate and on a convex spherical substrate~\cite{Guzzardi2007}.  However, it is pointed out by Mechkov et al.~\cite{Mechkov2007} that this instability is unphysical when the molecular interaction near the three-phase contact line is included using the disjoining pressure by the interface-displacement model~\cite{Indekeu1992}. Also,  Guzzardi et al.~\cite{Guzzardi2006,Guzzardi2007} introduced a notion of residual stability that the size limit of droplet dictated from the magnitude of line tension should be smaller than the wavelength of the unstable mode.  We will leave this problem of fluctuation and inclusion of the disjoining pressure on a concave spherical substrate for future investigation.

So far, we have considered a hydrophobic wall with $\theta_{\rm Y}=120^{\circ}$.  The result for the hydrophilic wall can be obtained analogously since the hydrophobic wall ($\theta_{\rm Y}>\theta_{\rm c}=90^{\circ}$) and hydrophilic wall ($\theta_{\rm Y}<\theta_{\rm c}=90^{\circ}$) are symmetric (Fig.~\ref{fig:CC5} ).  On the hydrophilic wall, however, the drying transition on the hydrophobic wall will be replaced by the wetting transition from a lens-shaped droplet with concave meniscus (LCU) to a spherical bubble (BB). Then, the drying transition line $\tilde{\tau}_{\rm dry}\left(\theta_{\rm e}\right)$ should be replace by the transition line $\tilde{\tau}_{\rm wet}\left(\theta_{\rm e}\right)$ defined by
\begin{eqnarray}
\tilde{\tau}
&=&
\left(\left(1-v\right)^{2/3}-\cos\theta_{\rm Y}-\frac{\left(1+\rho_{\rm e}-\xi_{\rm e}\right)^{2}\left(\cos\theta_{\rm e}+1+\xi_{\rm e}\right)}
{4\xi_{\rm e}}\right) \nonumber \\
&/&\left(\frac{\left(1+\rho_{\rm e}\cos\theta_{\rm e}-\xi_{\rm e}\right)}{2\rho_{\rm e}\sin\theta_{\rm e}}\right)\equiv \tilde{\tau}_{\rm wet}\left(\theta_{\rm e}\right), \nonumber \\
&&(\theta_{\rm e}>\theta_{\infty}, \mbox{Convex}),
\label{eq:C28}
\end{eqnarray}
derived from $f_{\rm lens}=f_{\rm bubble}$.  

\begin{figure}[htbp]
\begin{center}
\includegraphics[width=0.70\linewidth]{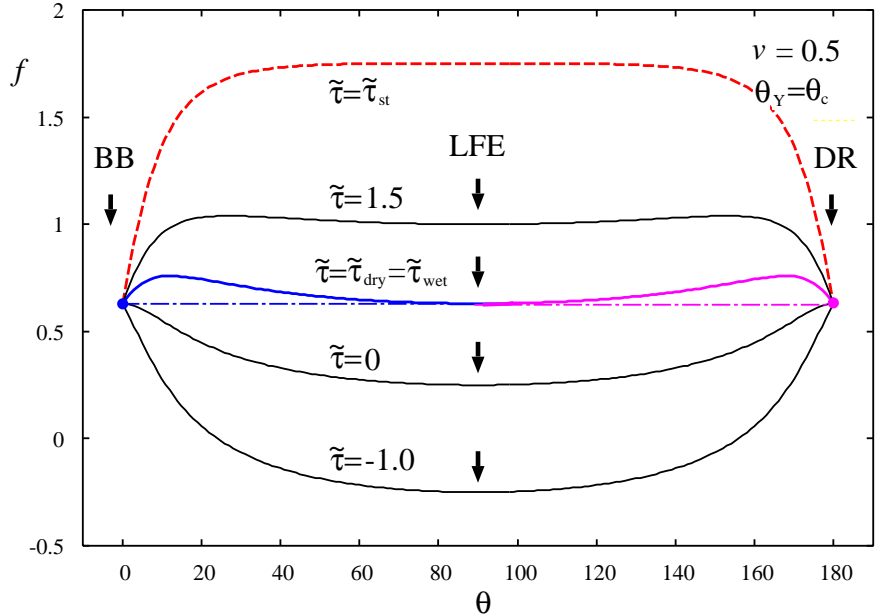}
\caption{
The Helmholtz free energy landscape of Eq.~(\ref{eq:C12}) when $\theta_{\rm Y}=\theta_{\infty}=\theta_{\rm c}=90^{\circ}$.  Now the contact angle of the lens-shaped droplet LFE is fixed at $\theta_{\rm c}=90^{\circ}$.  When $\tilde{\tau}=\tilde{\tau}_{\rm wet}=\tilde{\tau}_{\rm dry}\simeq 0.760$, the first-order wetting as well as drying transition take place simultaneously: a lens-shaped droplet may transform into either a spherical droplet or a spherical bubble.  When $\tilde{\tau}>\tilde{\tau}_{\rm st}\simeq 3.00$, a metastable lens-shaped droplet LFE becomes unstable.  Then, the only stable morphologies are either a spherical droplet (DR) or a spherical bubble (BB). 
  }
\label{fig:CC7}
\end{center}
\end{figure}

When $\theta_{\rm Y}=\theta_{\rm c}=90^{\circ}$, the equilibrium contact angle $\theta_{\rm e}$ is fixed at $\theta_{\rm c}$ as shown in Fig.~\ref{fig:CC7} since $\theta_{\infty}=90^{\circ}$ and the contact line coincides with the equator.  The lens-shaped droplet LFE has a flat meniscus as shown in Fig.~\ref{fig:CC1}(b) since $v=0.5$.   Furthermore, since $\theta_{\rm Y}=\theta_{\rm Y,w}=90^{\circ}$ (Fig.~\ref{fig:CC4}) as defined by Eq.~(\ref{eq:C25}), the two curves which represents wetting and drying overlap ($\tilde{\tau}_{\rm wet}\left(\theta_{\rm e}\right)=\tilde{\tau}_{\rm dry}\left(\theta_{\rm e}\right)$) from Eqs.~(\ref{eq:C27}) and (\ref{eq:C28}), and the wetting transition and the drying transition will coexist at $\tilde{\tau}=\tilde{\tau}_{\rm dry}=\tilde{\tau}_{\rm wet}\simeq 0.760$ determined from Eqs. (\ref{eq:C27}), (\ref{eq:C28}) and (\ref{eq:C26}) as shown in Fig.~\ref{fig:CC7}.  The lens-shaped droplet with a flat meniscus will transform either into a spherical droplet or into a spherical bubble at the same magnitude of the line tension $\tilde{\tau}=\tilde{\tau}_{\rm dry}=\tilde{\tau}_{\rm wet}$.  The metastable lens-shaped droplet LFE will become unstable above $\tilde{\tau}\ge \tilde{\tau}_{\rm st}\simeq 3.00$.  Then, only a spherical droplet or a spherical bubble can exist, whose probability will be 50:50.

\begin{figure}[htbp]
\begin{center}
\includegraphics[width=0.70\linewidth]{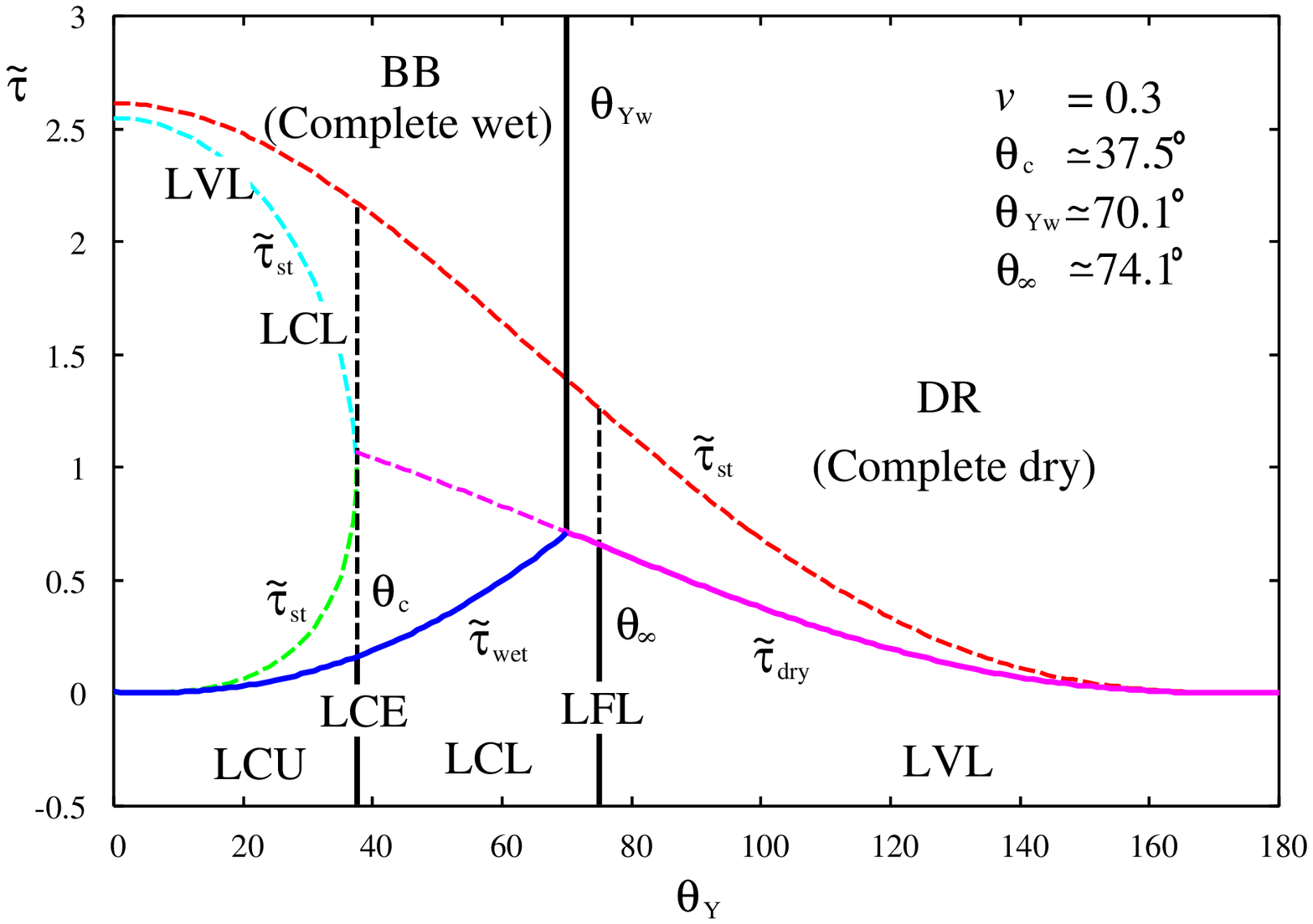}
\caption{
The morphological phase diagram of the droplet placed on the bottom of a spherical cavity when $v=0.3$ on the $\theta_{\rm Y}-\tilde{\tau}$ plane.  The phase diagram is more complex than that for $v=0.5$ in Fig.~\ref{fig:CC5}.
  }
\label{fig:CC8}
\end{center}
\end{figure}

Figure \ref{fig:CC8} shows the morphological phase diagram when $v=0.3$.  In this case $\theta_{\rm c}\simeq 37.5^{\circ}$, $\theta_{\rm Y,w}\simeq 70.1^{\circ}$ and $\theta_{\infty}\simeq 74.1^{\circ}$.  The phase diagram is divided into 5 regions by 5 boundaries, $\tilde{\tau}_{\rm wet}$, $\tilde{\tau}_{\rm dry}$, $\theta=\theta_{\rm c}$, $\theta=\theta_{\rm Y,w}$ and $\theta=\theta_{\infty}$.  In these five regions, the most stable morphologies are LCU, LCL, LVL, BB and DR (Fig.~\ref{fig:CC2}).  The boundary between LCU and LCL is LCE shown in Fig.~\ref{fig:CC1}(a).  The boundary between LCL and LVL is LFL, shown in Fig.~\ref{fig:CC2}(g).  In addition to these stable structures, there appears a metastable lens-shaped droplet which leads to the complex phase diagram in Fig.~\ref{fig:CC8}.

\begin{figure}[htbp]
\begin{center}
\subfigure[]
{
\includegraphics[width=0.45\linewidth]{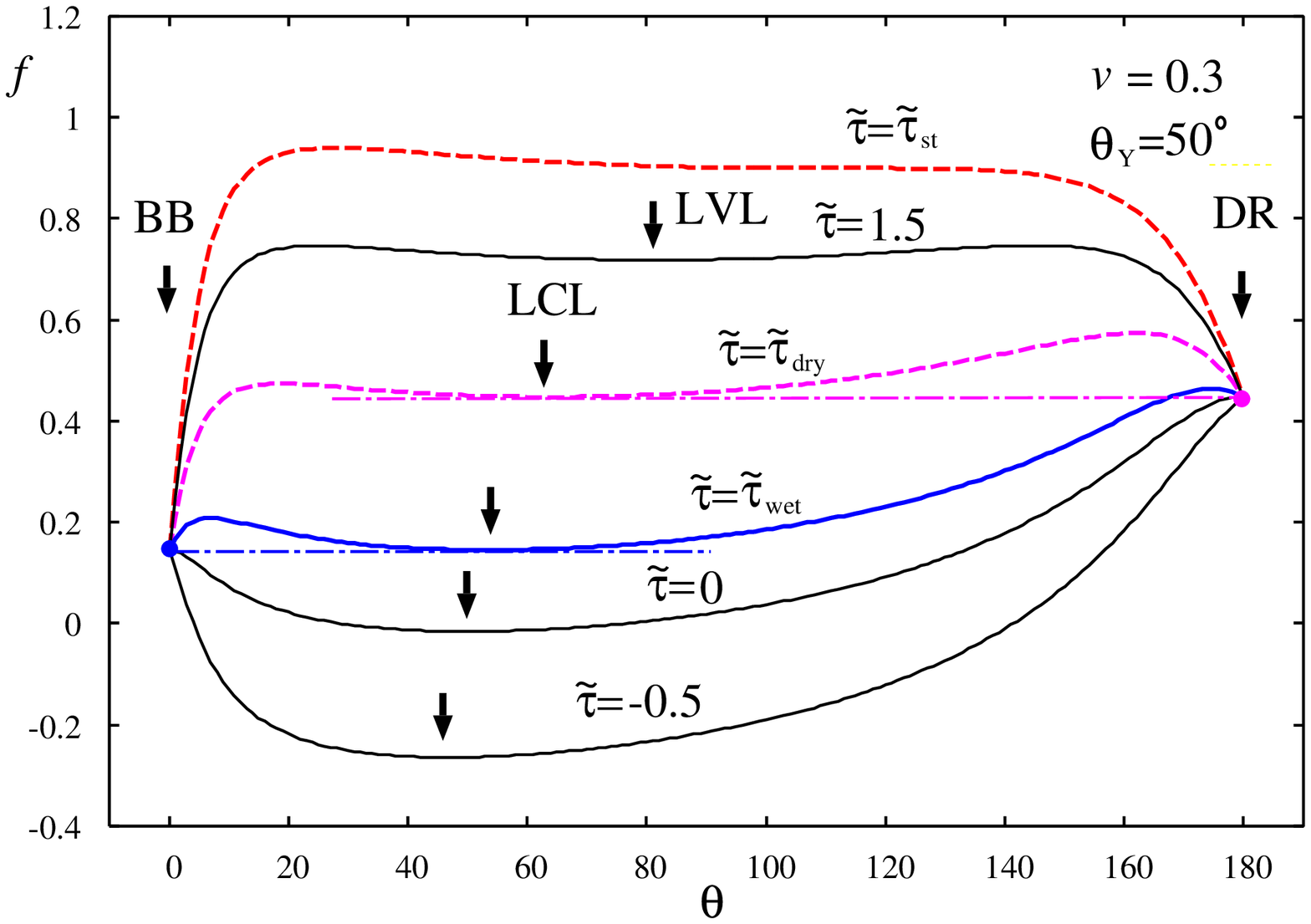}
\label{fig:9a}
}
\subfigure[]
{
\includegraphics[width=0.45\linewidth]{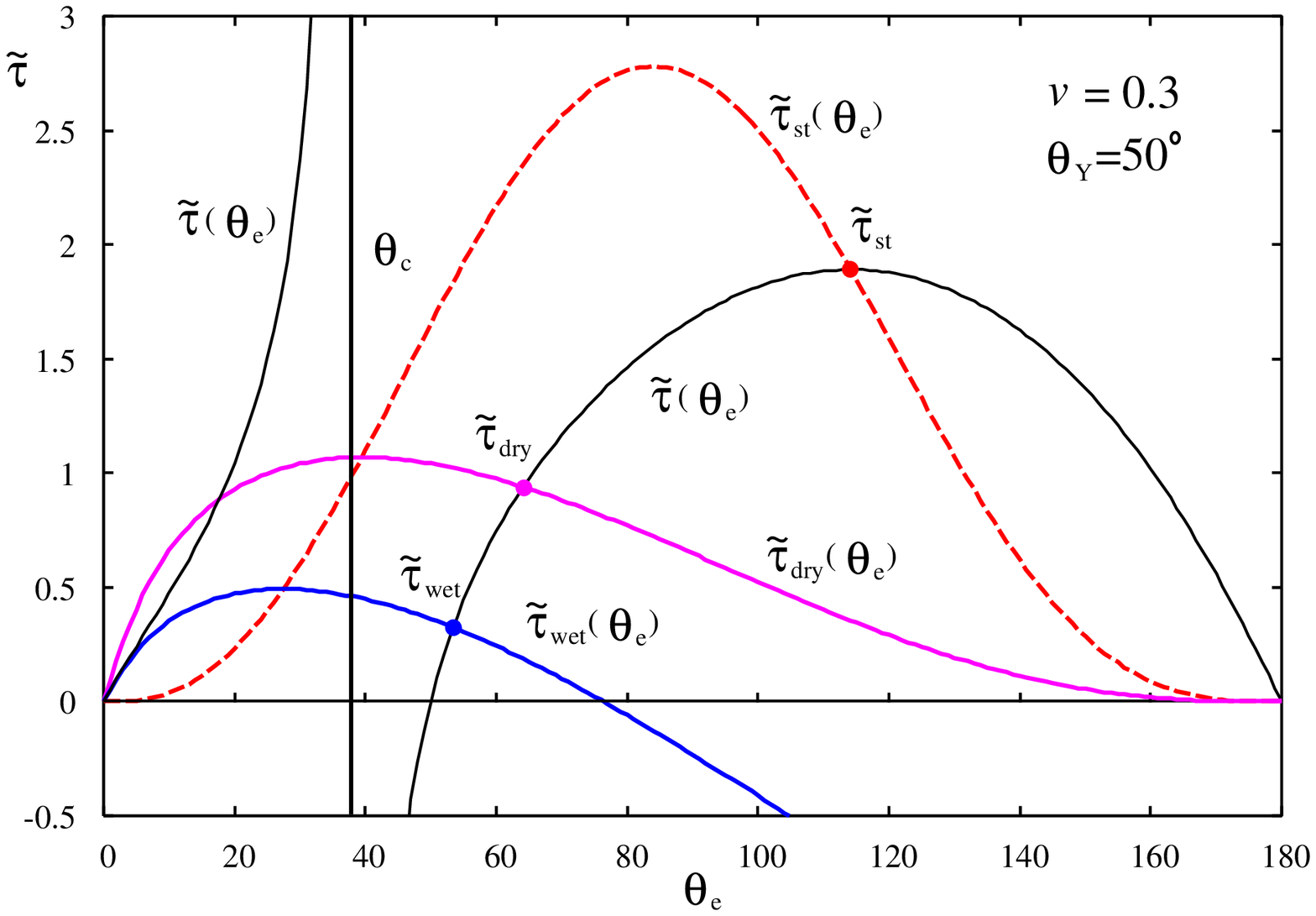}
\label{fig:9b}
}
\end{center}
\caption{
(a) The Helmholtz free energy landscape of Eq.~(\ref{eq:C12}) when $\theta_{\rm Y}=50^{\circ}$.  When $\tilde{\tau}=0$, the equilibrium contact angle $\theta_{\rm e}$ indicated by an arrow is given by the bare Young's contact angle $\theta_{\rm e}=\theta_{\rm Y}=50^{\circ}$.  It increases as the line tension $\tilde{\tau}$ is increased.  When $\tilde{\tau}=\tilde{\tau}_{\rm wet}\simeq 0.325$, the first-order wetting transition takes place: a lens-shaped droplet LCL will transform into a spherical bubble BB.  When $\tilde{\tau}=\tilde{\tau}_{\rm dry}\simeq 0.939$, a metastable lens-shaped droplet LCL may transform into a metastable spherical droplet DR rather than a bubble BB.   This metastable LCL transforms into LVL at $\tilde{\tau}_{\infty}\simeq 1.30$ which corresponds to $\theta_{\rm e}=\theta_{\infty}\simeq74.1^{\circ}$.  When the line tension reaches $\tilde{\tau}=\tilde{\tau}_{\rm st}\simeq 1.891$, a lens-shaped droplet LVL finally becomes unstable.  Then, only a stable bubble BB and a metastable droplet DR can exist.  
 (b) The determination of the equilibrium contact angle $\theta_{\rm e}$ as a function of the line tension $\tilde{\tau}$.  The three transition points $\tilde{\tau}_{\rm wet}\simeq 0.325$, $\tilde{\tau}_{\rm dry}\simeq 0.939$ and $\tilde{\tau}_{\rm st}\simeq 1.891$ are determined from the intersections of the four curves, $\tilde{\tau}\left(\theta_{\rm e}\right)$,  $\tilde{\tau}_{\rm wet}\left(\theta_{\rm e}\right)$, $\tilde{\tau}_{\rm dry}\left(\theta_{\rm e}\right)$  and  $\tilde{\tau}_{\rm st}\left(\theta_{\rm e}\right)$. 
 } 
\label{fig:CC9}
\end{figure}

Figure \ref{fig:CC9}(a) shows the free energy landscape of Eq~(\ref{eq:C12}) for a hydrophilic wall with $\theta_{\rm Y}=50^{^\circ}$ as a function of the contact angle $\theta$.  In this case, not only the wetting transition from a stable lens-shaped droplet LCL to a stable spherical bubble BB at $\tilde{\tau}_{\rm wet}\simeq 0.325$ but also the drying transition from a metastable lens-shaped droplet LCL to a metastable spherical droplet DR at $\tilde{\tau}=\tilde{\tau}_{\rm dry}\simeq 0.939$ may take place (Fig.~\ref{fig:CC8}).  There are three free energy extrema, which correspond to a spherical bubble BB, a lens-shaped droplet LCL and a spherical droplet DR.  The free energy minimum for LCL is the lowest as far as $\tilde{\tau}<\tilde{\tau}_{\rm wet}\simeq 0.325$.  At $\tilde{\tau}=\tilde{\tau}_{\rm wet}$, the two minima that correspond to LCL and BB have the same free energy so that they can coexist.  Then the {\it wetting transition} of the cavity wall occurs and the droplet spread over the whole surface to leave a spherical bubble (Fig.~\ref{fig:CC2}(b)).  When $\tilde{\tau}>\tilde{\tau}_{\rm wet}$, the LCL becomes metastable and the most stable morphology is BB. However, when the line tension reaches $\tilde{\tau}=\tilde{\tau}_{\rm dry}\simeq 0.939$, a metastable lens-shaped droplet LCL coexists with a metastable spherical droplet DR.  Then this metastable lens-shaped droplet LCL may transform into the metastable spherical droplet DR rather than the stable spherical bubble BB, thought the probability of transformation to droplet should be smaller than that to the bubble.  Then the transition between metastable states similar to drying transition could occur.  Finally, when $\tilde{\tau}>\tilde{\tau}_{\rm st}\simeq 1.891$, the lens shaped droplet becomes unstable.  Then only a stable spherical bubble BB and a metastable spherical droplet DR will survive.

Figure \ref{fig:CC9}(b) shows the determination of the equilibrium contact angle $\theta_{\rm e}$ and the three line tensions $\tilde{\tau}_{\rm wet}\simeq 0.325$, $\tilde{\tau}_{\rm dry}\simeq 0.939$, $\tilde{\tau}_{\rm st}\simeq 1.891$ and their corresponding equilibrium contact angle $\theta_{\rm e}$ when $\theta_{\rm Y}=50^{\circ}$.  The line tension $\tilde{\tau}_{\rm wet}$ that corresponds to the wetting transition is determined from the intersection of two curves $\tilde{\tau}\left(\theta_{\rm e}\right)$ given by Eq.~(\ref{eq:C26}) and  $\tilde{\tau}_{\rm wet}\left(\theta_{\rm e}\right)$ given by Eq.~(\ref{eq:C28}).  The drying transition $\tilde{\tau}_{\rm dry}$ is determined from the intersection of two curves $\tilde{\tau}\left(\theta_{\rm e}\right)$ and  $\tilde{\tau}_{\rm dry}\left(\theta_{\rm e}\right)$.  Finally, the stability limit $\tilde{\tau}_{\rm st}$ of a lens-shaped droplet is determined from the intersection of two curves $\tilde{\tau}\left(\theta_{\rm e}\right)$ and  $\tilde{\tau}_{\rm st}\left(\theta_{\rm e}\right)$.  The equilibrium contact angle $\theta_{\rm e}$ increases as the line tension $\tilde{\tau}$ is increased (see also Fig.~\ref{fig:CC9}(a)).  The metastable lens-shaped droplet LCL for $\tilde{\tau}_{\rm st}\ge \tilde{\tau}\ge \tilde{\tau}_{\rm wet}$ will change its morphology from LCL to LVL via LEL when $\theta_{\rm e}=\theta_{\infty}\simeq 74.1^{\circ}$ and $\tilde{\tau}_{\infty}\simeq 1.30$.

\begin{figure}[htbp]
\begin{center}
\includegraphics[width=0.70\linewidth]{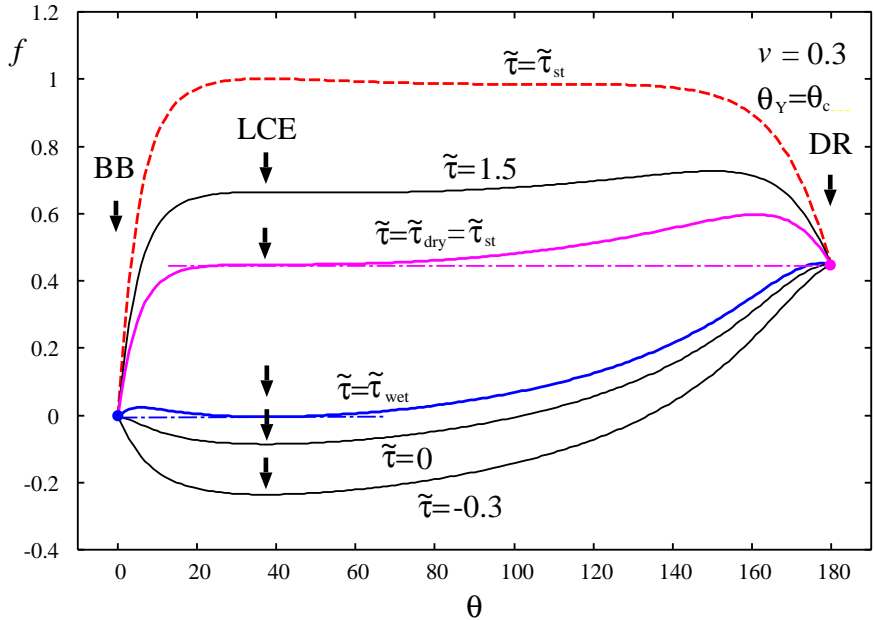}
\caption{
The Helmholtz free energy landscape of Eq.~(\ref{eq:C12}) when $\theta_{\rm Y}=\theta_{\rm c}\simeq 37.5^{\circ}$.  Now the contact angle of the lens-shaped droplet is fixed at $\theta_{\rm c}\simeq 37.5^{\circ}$, and the contact line is also fixed at the equator.  When $\tilde{\tau}=\tilde{\tau}_{\rm wet}\simeq 0.162$, the first-order wetting transition takes place: a lens-shaped droplet may transform into a spherical bubble.  When $\tilde{\tau}=\tilde{\tau}_{\rm dry}\simeq 1.068$, a metastable lens-shaped droplet becomes unstable as $\tilde{\tau}_{\rm dry}=\tilde{\tau}_{\rm st}$.  The lens-shaped droplet recover metastability again when $\tilde{\tau}>\tilde{\tau}_{\rm dry}$ (e.g. $\tilde{\tau}=1.5$).  This lens-shaped droplet finally becomes unstable when $\tilde{\tau}>\tilde{\tau}_{\rm st}\simeq 2.173$. 
 }
\label{fig:CC10}
\end{center}
\end{figure}

When Young's contact angle is equal to the characteristic contact angle $\theta_{\rm Y}=\theta_{\rm c}\simeq 37.5^{\circ}$ determined from Eq.~(\ref{eq:C18}), the equilibrium contact angle $\theta_{\rm e}$ is fixed at $\theta_{\rm c}$ and the contact line is fixed at the equator with a concave meniscus (LCE, Fig.~\ref{eq:C1}(a)) .  The drying transition $\tilde{\tau}_{\rm dry}$ from a metastable lens-shaped droplet LCE to a metastable droplet DR coincides with the stability limit $\tilde{\tau}_{\rm st}$ (see also Fig.~\ref{fig:CC8}).   Fig.~\ref{fig:CC10} shows the free energy landscape of Eq~(\ref{eq:C12}) for $\theta_{\rm Y}=\theta_{\rm c}\simeq 37.5^{^\circ}$ as a function of the contact angle $\theta$.  In this case, the wetting transition from a lens-shaped droplet LCE to a spherical bubble BB occurs when $\tilde{\tau}=\tilde{\tau}_{\rm wet}\simeq 0.162$. However, the transition from a metastable lens-shaped droplet LCE to a metastable spherical droplet DR similar to the drying transition, which was observed when $\theta_{\rm Y}=50^{\circ}$ becomes critical (Fig.~\ref{fig:CC8}) since $\tilde{\tau}_{\rm dry}=\tilde{\tau}_{\rm st}\simeq 1.068$.  This pseudo-drying transition coincides with the stability limit and it cannot occur.  Above this critical point ($\tilde{\tau}>\tilde{\tau}_{\rm dry}=\tilde{\tau}_{\rm st}$), a lens-shaped droplet can exist again as a metastable droplet (see Fig.~\ref{fig:CC10} with $\tilde{\tau}=1.5$).  It becomes unstable at the second stability limit $\tilde{\tau}_{\rm st}=2.173$.  The lens-shaped droplet becomes unstable at an isolated point $\tilde{\tau}=\tilde{\tau}_{\rm st}\simeq 1.068$ and above the second stability limit $\tilde{\tau}\ge \tilde{\tau}_{\rm st}\simeq 2.173$. During the course of these morphological transformation, the equilibrium contact angle $\theta_{\rm e}$ for the lens-shaped droplet indicated by an arrow is always fixed at $\theta_{\rm c}=37.5^{\circ}$ and the contact line is fixed at the equator.

When $\theta_{\rm Y}<\theta_{\rm c}$, the phase diagram is more complex (Fig.~\ref{fig:CC8}).  A lens-shaped droplet with a concave meniscus whose contact line is located on the upper hemisphere (LCU) will transform into a spherical bubble BB at $\tilde{\tau}=\tilde{\tau}_{\rm wet}$.  Above $\tilde{\tau}_{\rm wet}$, this lens-shaped droplet LCU will be metastable.  It will be unstable at the first (lowest) stability limit $\tilde{\tau}_{\rm st}$.  In fact, there appear three stability limit, the first (lowest), the second, and the last (highest) stability limits (Fig.~\ref{fig:CC8}).  In the region between the lowest stability limit and the second stability limit, a lens-shaped droplet of any morphology cannot exist, even as a metastable droplet.  A metastable lens-shaped droplet reappears above the second stability limit.  However, the morphology becomes a lens-shaped droplet with concave meniscus whose contact line locates on the lower hemisphere (LCL).  The contact line jumps from the upper hemisphere of LCU below the first stability limit to the lower hemisphere of LCL above the second stability limit.  This LCL changes morphology to LVL via LEL at $\theta_{\rm e}=\theta_{\infty}$, and finally becomes unstable at the third (highest) stability limit $\tilde{\tau}_{\rm st}$.

\begin{figure}[htbp]
\begin{center}
\subfigure[]
{
\includegraphics[width=0.45\linewidth]{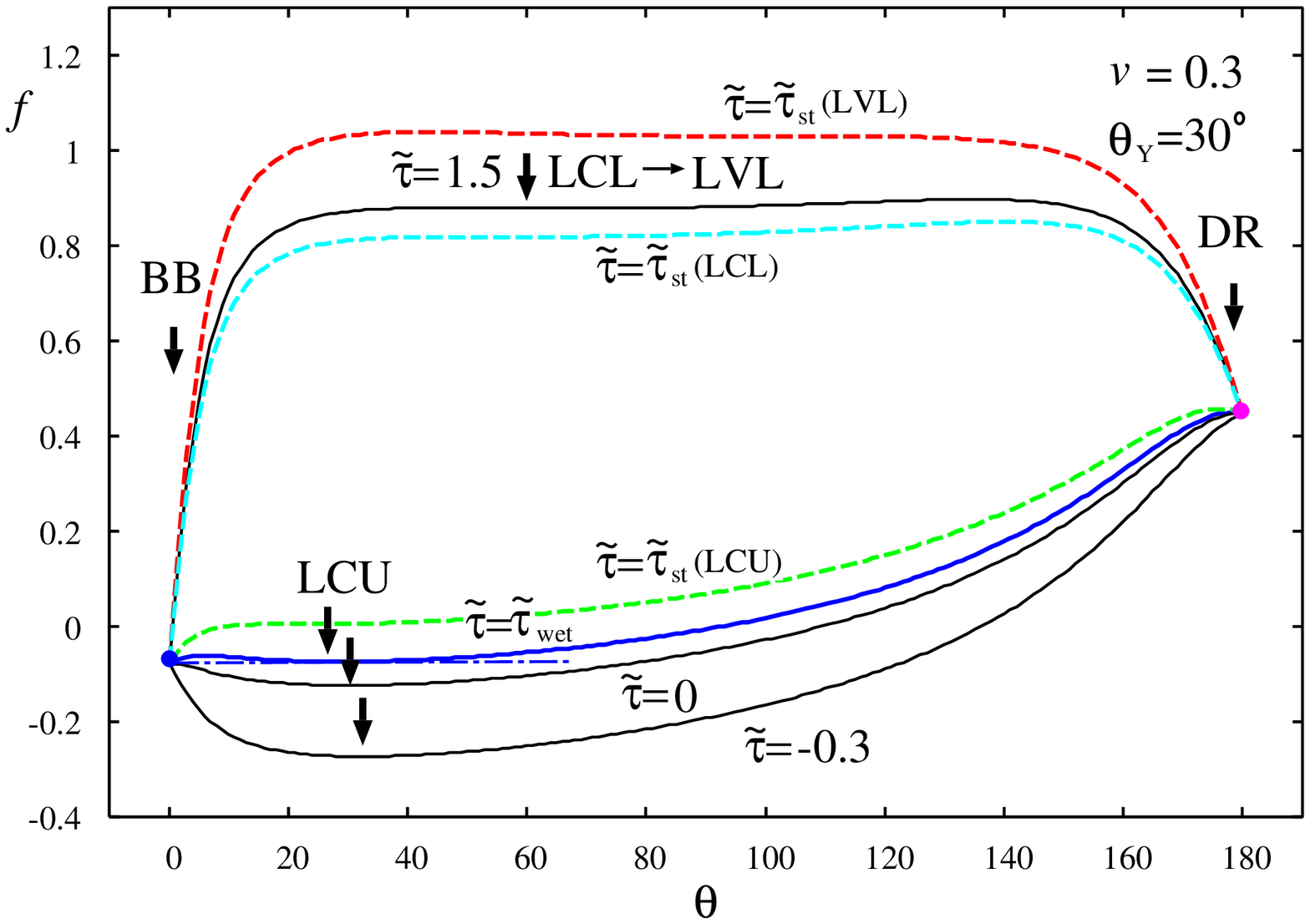}
\label{fig:11a}
}
\subfigure[]
{
\includegraphics[width=0.45\linewidth]{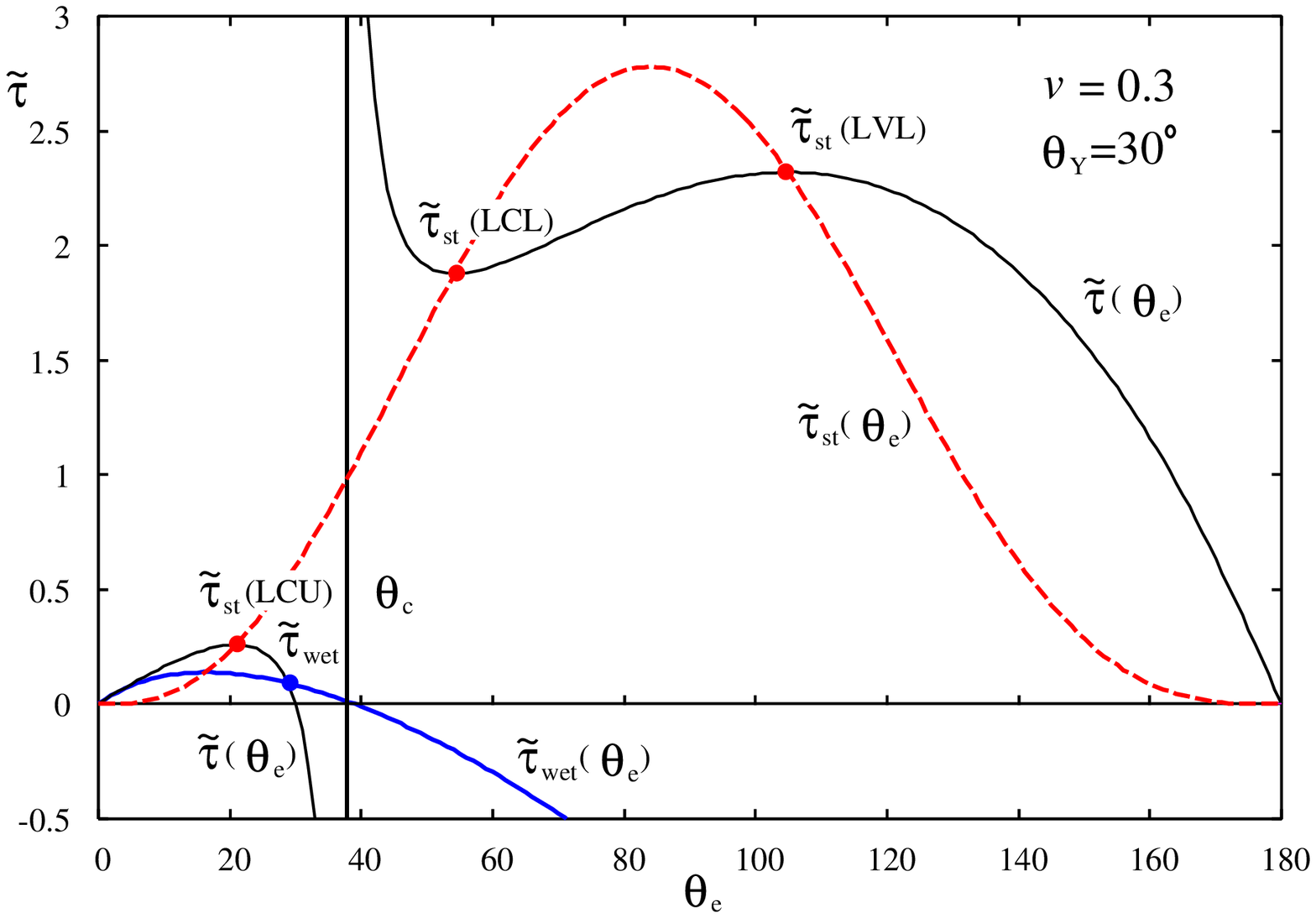}
\label{fig:11b}
}
\end{center}
\caption{
(a) The Helmholtz free energy landscape of Eq.~(\ref{eq:C12}) when $\theta_{\rm Y}=30^{\circ}$.  When $\tilde{\tau}=0$, the contact angle is given by the bare Young's contact angle $\theta=\theta_{\rm Y}=30^{\circ}$.  It decreases as the line tension $\tilde{\tau}$ is increased.  When $\tilde{\tau}=\tilde{\tau}_{\rm wet}\simeq 0.100$, the first-order wetting transition takes place: a lens-shaped droplet LCU will transform into a spherical bubble BB.  Above the first stability limit $\tilde{\tau}>\tilde{\tau}_{\rm st}\simeq 0.260$, a metastable lens-shaped droplet LCU will be unstable and cannot appear until the line tension reaches the second stability limit $\tilde{\tau}_{\rm st}\simeq 1.878$, where the metastable LCL rather than LCU appears.  By increasing the line tension further, this LCL changes its morphology to LVL, which will be unstable when the line tension finally reaches the third (highest) stability limit $\tilde{\tau}_{\rm st}\simeq 2.319$.
 (b) The determination of the equilibrium contact angle $\theta_{\rm e}$ as a function of the line tension $\tilde{\tau}$.  The wetting transition points $\tilde{\tau}_{\rm wet}\simeq 0.100$ is determined from the intersection of the curve  $\tilde{\tau}\left(\theta_{\rm e}\right)$ and $\tilde{\tau}_{\rm wet}\left(\theta_{\rm e}\right)$.  The three stability limits  $\tilde{\tau}_{\rm st}\simeq 0.290$,  $\tilde{\tau}_{\rm st}\simeq 1.878$ and  $\tilde{\tau}_{\rm st}\simeq 2.319$ are determined from the intersection of $\tilde{\tau}\left(\theta_{\rm e}\right)$ and  $\tilde{\tau}_{\rm st}\left(\theta_{\rm e}\right)$. Note that the contact angle of the flat meniscus and, therefore, the LCL-LVL boundary is at $\theta_{\rm e}=\theta_{\infty}\simeq 74.1^{\circ}$ and $\tilde{\tau}_{\infty}\simeq 2.08$.
 } 
\label{fig:CC11}
\end{figure}

Figure \ref{fig:CC11}(a) shows the free energy landscape of Eq~(\ref{eq:C12}) for a hydrophilic wall with $\theta_{\rm Y}=30^{^\circ}$ as a function of the contact angle $\theta$.  In this case, only the wetting transition from a lens-shaped droplet LCU to a spherical bubble BB will be observed (Fig.~\ref{fig:CC8}).  The shallow free energy minimum for LCU is the lowest as far as $\tilde{\tau}<\tilde{\tau}_{\rm wet}\simeq 0.100$.  The equilibrium contact angle $\theta_{\rm e}$ indicated by an arrow decreases as the line tension is increased.  At $\tilde{\tau}=\tilde{\tau}_{\rm wet}$, the two minima that correspond to LCU and BB have the same free energy so that they can coexist.  Then the {\it wetting transition} from a lens-shaped droplet LCU to a spherical bubble BB (Fig.~\ref{fig:CC2}(b)) occurs.

When $\tilde{\tau}>\tilde{\tau}_{\rm wet}$, the LCU becomes metastable.  This metastable lens-shaped droplet LCU becomes unstable when the line tension reaches the first (lowest) stability limit $\tilde{\tau}_{\rm st}\simeq 0.260$.  When the line tension is between the lowest stability limit $\tilde{\tau}_{\rm st}\simeq 0.260$ and the second stability limit $\tilde{\tau}_{\rm st}\simeq 1.878$, the lens-shaped droplet become unstable and cannot exist.  Above the second stability limit, the lens-shaped droplet becomes metastable again.  However, the morphology changes from LCU to LCL.  This lens-shaped droplet LCL changes it morphology to LVL via LEL at $\theta_{\rm e}=\theta_{\infty}\simeq 74.1^{\circ}$, and it finally becomes unstable at the last (highest) stability limit $\tilde{\tau}_{\rm st}\simeq 2.319$.

\begin{figure}[htbp]
\begin{center}
\includegraphics[width=0.70\linewidth]{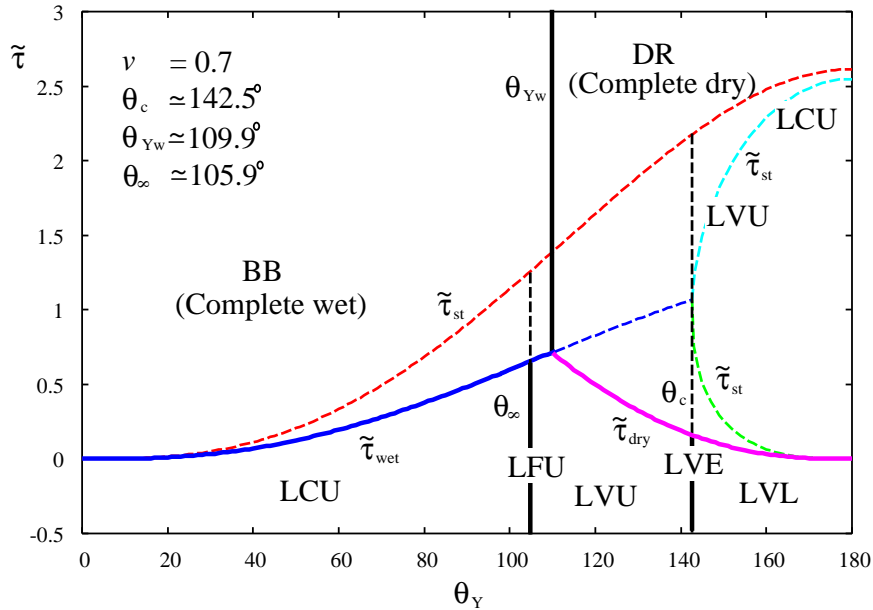}
\caption{
The morphological phase diagram of a droplet placed on the bottom of a spherical cavity when $v=0.7$ on the $\theta_{\rm Y}-\tilde{\tau}$ plane. The phase diagram is a reflection image of Fig.~\ref{fig:CC8} about $\theta_{\rm Y}=90^{\circ}$.  
 }
\label{fig:CC12}
\end{center}
\end{figure}

Figure \ref{fig:CC11}(b) shows the determination of $\tilde{\tau}_{\rm wet}$ of the wetting transition and three stability limits $\tilde{\tau}_{\rm st}$.  The line tension $\tilde{\tau}_{\rm wet}\simeq 0.100$ that corresponds to the wetting transition is determined from the intersection of two curves $\tilde{\tau}\left(\theta_{\rm e}\right)$ given by Eq.~(\ref{eq:C26}) and  $\tilde{\tau}_{\rm wet}\left(\theta_{\rm e}\right)$ given by Eq.~(\ref{eq:C28}).  The stability limits $\tilde{\tau}_{\rm st}\simeq 0.260, 1.878$ and $2.319$ of a lens-shaped droplet are determined from the intersection of two curves $\tilde{\tau}\left(\theta_{\rm e}\right)$ given by Eq.~(\ref{eq:C26}) and  $\tilde{\tau}_{\rm st}\left(\theta_{\rm e}\right)$ given by Eq.~(\ref{eq:C17}).  The lowest root $\tilde{\tau}_{\rm st}\simeq 0.260$ corresponds to the stability limit for the lens-shaped droplet with concave meniscus when the contact line is on the upper hemisphere (LCU) as the equilibrium contact angle satisfies $\theta_{\rm e}<\theta_{\rm c}<\theta_{\infty}$.  The contact angle of this LCU decreases as the line tension is increased.  On the other hand, the second stability limit $\tilde{\tau}_{\rm st}\simeq 1.878$ is the lower stability limit of the lens shaped droplet above which the lens-shaped droplet LCL with concave meniscus whose contact line is located on the lower hemisphere ($\theta_{\rm c}<\theta_{\rm e}<\theta_{\infty}$) appears as a metastable droplet.  The contact angle of this LCL increases as the line tension is increased.  The meniscus changes from concave to convex so that the droplet morphology changes from LCL to LVL via LEL at $\theta_{\rm e}=\theta_{\infty}\simeq 74.1^{\circ}$ and $\tilde{\tau}_{\infty}\simeq 2.08$. The contact angle continue to increase as the line tension is increased until it reaches the third (highest) stability limit $\tilde{\tau}_{\rm st}\simeq 2.319$ where LVL becomes unstable.

When the volume of the droplet is larger than $v=0.5$, the droplet and the bubble exchange their roles (Fig.~\ref{fig:CC2}).  The morphological phase diagram of $v=0.7$ shown in Fig.~\ref{fig:CC12} is a reflection image of Fig.~\ref{fig:CC8} for $v=0.3$ about $\theta_{\rm Y}=90^{\circ}$.  The hydrophilicity and hydrophobicity, and, therefore, the droplet and the bubble exchange their roles between $v=0.3$ and $v=0.7$.  Therefore, the scenario of morphological transition for $v=0.7$ in Fig.~\ref{fig:CC12} can be understood from that for $v=0.3$ in Fig.~\ref{fig:CC8} by exchanging the role of wetting and drying, as well as those of the droplet and the bubble.

\section{\label{sec:sec5}Conclusion}

In this study, we considered the line-tension effects on the morphology of a lens-shaped droplet and bubble of fixed volume placed on the inner wall of a spherical cavity within the capillary model. The contact angle is determined from the generalized Young's equation, which takes into account the effects of line-tension.  The morphology is studied using the obtained mathematically rigorous formula for the Helmholtz free energy.  Not only the morphological transition known as the drying transition from a lens-shaped droplet to a spherical droplet but also that known as wetting transition from a lens-shaped droplet to a wetting layer which leads to a spherical bubble are predicted. The scenarios of these morphological transitions were examined using the free-energy landscape of the Helmholtz free energy.  In addition to these first-order like morphological transitions between thermodynamically {\it stable} morphology, the morphological transition between a {\it metastable} lens-shaped droplet, the {\it metastable} spherical droplet, and the {\it metastable} spherical bubble was found. Therefore, the phase diagram of the morphological transition in a cavity is more complex than that on a flat substrate, although the system considered is still relatively simple.

In addition, we found a special role played by the equator of the spherical cavity, where the contact-line length of a droplet reaches its maximum, which was already found for the droplet on a convex spherical substrate.  The contact line of the droplet cannot cross the equator by continuously changing the magnitude of the line tension.  When the contact line coincides with the equator or the contact angle is given by the characteristic contact angle,  the contact line of the droplet is fixed at the equator and cannot move.  In this special circumstance, the droplet is a special droplet which cannot change contact angle continuously.

In conclusion, we studied various scenarios of the morphological change of a lens-shaped droplet and bubble placed on the inner wall of a spherical cavity using a mathematically rigorous formula for the Helmholtz free energy.  We used the simplest capillary model, and neglected the non-circular fluctuation of the contact line.  The former will be important if the liquid-substrate interaction is long-ranged represented by the disjoining pressure, and the latter will cause the instability of the lens-shaped droplet if the line tension is negative.  The problem of the fluctuation of the contact line and the effect of the disjoining pressure are left for future investigation.

\begin{acknowledgments}
A part of this study was conducted during the author’s visit to School of Physics, Universiti Sains Malaysia (USM, Penang, Malaysia) and Faculty of Engineering and Technology, Multimedia University (MMU, Malacca, Malaysia).  The author is grateful to Dr. T. L. Yoon (USM) and Dr. T. L. Lim (MMU) for their helpful discussions and warm hospitalities. 
\end{acknowledgments}

\appendix*
\section{Derivation of various analytical formulas}
\begin{figure}[htbp]
\begin{center}
\includegraphics[width=0.80\linewidth]{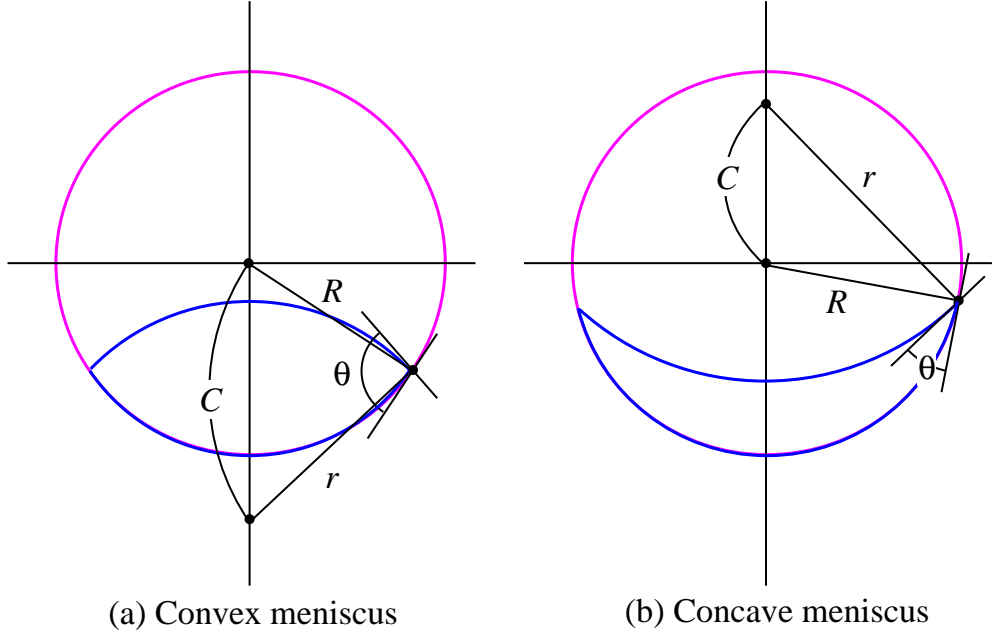}
\caption{
(a) A lens-shaped droplet with convex meniscus. (b) A lens-shaped droplet with concave meniscus.  Instead of using the contact angle $\theta$ as one of the independent variables, we use the center to center distance $C$ between two spheres in addition to the two radius $R$ and $r$. 
   }
\label{fig:A1}
\end{center}
\end{figure}

Here, we sketch a mathematical technique to derive various analytical formulae such as Eqs. (5) and (12) in the main text.  In short, the derivation is based on the integration scheme proposed by Hamaker~\cite{Hamaker1937} and a change of variable~\cite{Iwamatsu2015a,Iwamatsu2015b} from the contact angle $\theta$ to the distance $C$ between the centers of two spheres of the substrate (cavity) and the droplet shown in Fig.~\ref{fig:A1}.  By using this simple variable $C$, we can avoid tedious and complicated transformation of trigonometric functions.  Since all equations are linear in  $\sigma_{\rm sv}$, $\sigma_{\rm sl}$, $\sigma_{\rm lv}$ and $\tau$, manipulation is tedious but straightforward.  

Using the variables $R$, $r$ and $C$, the Helmholtz free energy of the droplet is given by
\begin{equation}
F = \pi r \sigma_{\rm lv}\frac{R^{2}-\left(C-r\right)^{2}}{C} + \pi R \Delta\sigma\frac{r^2-\left(C-R\right)^{2}}{C}+2\pi \tau\frac{rR}{C}\sin\theta,\;\;\;\;(\mbox{Convex})
\label{eq:AA1}
\end{equation}
for a convex meniscus and
\begin{equation}
F = \pi r \sigma_{\rm lv}\frac{R^{2}-\left(C-r\right)^{2}}{C} - \pi R \Delta\sigma\frac{r^2-\left(C+R\right)^{2}}{C}+2\pi \tau\frac{rR}{C}\sin\theta,\;\;\;\;(\mbox{Concave})
\label{eq:AA2}
\end{equation}
for a concave meniscus.  Minimizing these free energy $F$ under the constraint of constant volume $V$ given by
\begin{equation}
V= \frac{\pi}{12C}\left(R+r-C\right)^{2}\left(C^{2}-3\left(R-r\right)^{2}+2C\left(R+r\right)\right),\;\;\;\;(\mbox{Convex})
\label{eq:AA3}
\end{equation}
and 
\begin{equation}
V=-\frac{\pi}{12C}\left(R-r+C\right)^{2}\left(C^{2}-3\left(R+r\right)^{2}-2C\left(R-r\right)\right),\;\;\;\;(\mbox{Concave}),
\label{eq:AA4}
\end{equation}
we obtain the equation which will determine the equilibrium contact angle.  Since, the equation for the concave meniscus will be obtained by changing the sign of radius (curvature) according to
\begin{equation}
C\rightarrow -C,\;\;\;\;r\rightarrow -r,
\label{eq:AA5}
\end{equation}
we will only present the results for the convex meniscus.  The results for the concave meniscus is easily obtained by the transformation given by Eq.~(\ref{eq:AA5}).  By minimizing the Helmholtz free energy, we can obtain the equation for the equilibrium contact angle $\theta_{\rm e}$ given by
\begin{equation}
\Delta\sigma=\frac{-C^{2}+R^{2}+r^{2}}{2rR}\sigma_{\rm lv}-\frac{C^{2}+R^{2}-r^{2}}{R\sqrt{-C^{4}-\left(R^{2}-r^{2}\right)^{2}+2C^{2}\left(R^{2}+r^{2}\right)}}\tau,\;\;\;\;(\mbox{Convex})
\label{eq:AA6}
\end{equation}
which will reduce to the generalized Young's equation given by Eq.~(14) of the main text and the minimum free energy of a lens-shaped droplet 
\begin{eqnarray}
F_{\rm lens}&=&\frac{\pi\left(-C+R+r\right)^{2}\left(C^{2}-\left(R-r\right)^{2}+2Cr\right)}{2Cr}\sigma_{\rm lv}
\nonumber \\
&-&\frac{2\pi\left(-C+R-r\right)\left(-C+R+r\right)}{\left(C+R-r\right)\left(C-R+r\right)\left(-C+R+r\right)\left(C+R+r\right)}\tau,\;\;\;\;(\mbox{Convex}) \nonumber \\
\label{eq:AA7}
\end{eqnarray}
which will reduces to Eqs.~(15) and (16) of the main text.  The results for the concave meniscus can be easily obtained by analogously using the transformation in Eq.~(\ref{eq:AA5}).

By differentiating Eq.~(\ref{eq:AA1}) with respect $r$ twice, we obtain
\begin{eqnarray}
\frac{d^{2}F}{dr^{2}}&=&
\frac{\pi\left(C^{2}-R^{2}-2Cr+r^{2}\right)^{2}\left(C^{2}-R^{2}+4Cr+r^{2}\right)}{Cr\left(C^{2}-R^{2}+2Cr+r^{2}\right)^{2}}\sigma_{\rm lv}
\nonumber \\
&+&\frac{16\pi Cr^{2}R^{2}\left(C^{2}-R^{2}-2Cr+r^{2}\right)}{\left(C^{2}-R^{2}+2Cr+r^{2}\right)^{3}\sqrt{-C^{4}-\left(R^{2}-r^{2}\right)+2C^{2}\left(R^{2}+r^{2}\right)}}\tau,\;\;\;\;(\mbox{Convex}) \nonumber \\
\label{eq:AA8}
\end{eqnarray}
The stability condition $d^{2}F/dr^{2}\ge 0$ leads to the Eq.~(17) of the main text.


\begin{thebibliography}{99}
\bibitem{Seemann2012} R. Seemann, M. Brinkmann, T. Pfohl and S. Herminghaus, Rep. Prog. Phys. {\bf 75}, 016601 (2012).
\bibitem{Lohse2015} D. Lohse and X. Zhang, Rev. Mod. Phys. {\bf 87}, 981 (2015).
\bibitem{Nosonovsky2007} M. Nosonovsky and B. Bhushan, Mater. Sci. Eng. R {\bf 58
}, 162 (2007).
\bibitem{Song2014} C. Song and Y. Zheng, J. Colloid Interface Sci. {\bf 427}, 2 (2014).
\bibitem{Gibbs1906} Gibbs, J. W. {\it The scientific papers of J. Willard Gibbs V1: Thermodynamics}, p. 288 footnote, Longmans and Green, London, 1906.
\bibitem{deGennes1985} P. G. de Gennes, Rev. Mod. Phys. {\bf 57}, 827 (1985).
\bibitem{Bonn2009} D. Bonn, J. Eggers, J. Indekeu,  J. Meunier, and E. Rolley, Rev. Mod. Phys. {\bf 81}, 739 (2009).
\bibitem{Weijs2011} J. H. Weijs, A. Marchand, B. Andreotti, D. Lohse, J. H. Snoeijer, Phys. Fluid. {\bf 23}, 022001 (2011).
\bibitem{Bormashenko2013} E. Yu. Bormashenko {\it Wetting of Real Surfaces}, De Gruyter, Berlin, 2013.
\bibitem{Pompe2000} T. Pompe and S. Herminghaus, Phys. Rev. Lett. {\bf 85}, 1930 (2000).
\bibitem{Wang2001} J. Y. Wang, S. Betelu, and B. M. Law, Phys. Rev. E {\bf 63}, 031601 (2001).
\bibitem{Checco2003} A. Checco, P. Guenoun, and J. Daillant, Phys. Rev. Lett. {\bf 91}, 186101 (2003). 
\bibitem{Schimmele2007} L. Schimmele, M. Napi\'orkowski and S. Dietrich, J. Chem. Phys. {\bf 127}, 164715 (2007).
\bibitem{Widom1995} B. Widom, J. Phys. Chem. {\bf 99}, 2803 (1995).
\bibitem{Navascues1981} G. Navascu\'es and P. Tarazona, J. Chem. Phys. {\bf 75}, 2441 (1981).
\bibitem{Singha2015} S. K. Singha, P. K. Das, and B. Maiti, J. Chem. Phys. {\bf 142}, 104706 (2015). 
\bibitem{Blecua2006} P. Blecua, R. Lipowsky, and J. Kierfeld, Langmuir {\bf 22}, 11041 (2006).
\bibitem{Guzzardi2007} L. Guzzardi and R. Rosso, J. Phys. A {\bf 40}, 19 (2007).
\bibitem{Hienola2007} A. I. Hienola, P. M. Winkler, P. E. Wagner, H. Venkam\"aki, A. Lauri, I. Napari and M. Kulmala, J. Chem. Phys. {\bf 126}, 094705 (2007).
\bibitem{Cooper2007} S. J. Cooper, C. E. Nichloson and J. A. Liu, J. Chem. Phys. {\bf 129}, 124715 (2008).
\bibitem{Dutka2010}  F. Dutka and M. Napi\'orkowski, J. Chem. Phys. {\bf 133}, 051101 (2010).
\bibitem{Iwamatsu2015a} M. Iwamatsu, Langmuir {\bf 31}, 3861 (2015).
\bibitem{Iwamatsu2015b} M. Iwamatsu, J. Chem. Phys. {\bf 143}, 014701 (2015).
\bibitem{Qiu2015} Y. Qiu and V. Molinero, J. Am. Chem. Soc. {\bf 137}, 10642 (2015).
\bibitem{Kubalsky2000} G. P. Kubalsky and M. Napi\'orkowski, J. Phys.: Condens. Matter {\bf 12}, 9221 (2000).
\bibitem{Dubrovskii2009} V. G. Dubrovskii, M. V. Nazarenko, N. V. Shibirev, Tech. Phys. Lett. {\bf 35}, 1117 (2009).
\bibitem{Extrand2012} C. W. Extrand and S. I. Moon, Langmuir {\bf 28}, 7775 (2012).
\bibitem{Maksimov2013} A. O. Maksimov, A. M. Kaverin, V. G. Baidakov, Langmuir {\bf 29}, 3924 (2013).
\bibitem{Maheshwari2016} S. Maheshwari, M. van der Hoef, and D. Lohse, Langmuir {\bf 32}, 316 (2016).
\bibitem{Ruckenstein2010} E. Ruckenstein and G. O. Berim, G. O. J. Colloid Interface Sci. {\bf 351}, 277 (2010).
\bibitem{Qian2012} M. Qian and J. Ma, J. Cryst. Growth {\bf 355}, 73 (2012).
\bibitem{Whyman2011} G. Whyman, E. Bormashenko, Langmuir {\bf 27}, 8171 (2011).
\bibitem{Tolman1949} R. C. Tolman, J. Chem. Phys. {\bf 17}, 333 (1949). 
\bibitem{Block2014} B. J. Block, D. Deb, F. Schmitz, A. Statt, A. Tr\"oster, A. Winkler, T. Zykova-Timan, P. Birnau, and K. Binder, Eur. Phys. J. Special Topics {\bf 223}, 347 (2014), and references therein. 
\bibitem{Young1805} T. Young, Phil. Trans. R. Soc. Lond. {\bf 95}, 65 (1805).
\bibitem{Hamaker1937} H. C. Hamaker, Physica {\bf 4}, 1058 (1937).
\bibitem{Drelich1996} J. Drelich, Coll. Surf. A {\bf 116}, 43 (1996).
\bibitem{Vandecan2008} Y. Vandecan and J. O. Indekeu, J. Chem. Phys. {\bf 128}, 104902 (2008).
\bibitem{Dobbs1999} H. Dobbs, Physica A {\bf 217}, 36 (1999). 
\bibitem{Brinkmann2005} M. Brinkmann, J. Kierfeld, and R. Lipowsky, J. Phys.: Condense Matter {\bf 17}, 2349 (2005).  
\bibitem{Guzzardi2006} L. Guzzardi, R. Rosso, E. G. Virga, Phys. Rev. E {\bf 73}, 021602 (2006). 
\bibitem{Mechkov2007} S. Mechkov, G. Oshanin, M. Rauscher, M. Brinkmann, A. M. Cazabat, and S. Dietrich, Europhys. Lett. {\bf 80}, 66002 (2007).
\bibitem{Indekeu1992}  J. O. Indekeu, Physica A {\bf 183}, 439 (1992). 
\end{thebibliography}

\end{document}